\documentclass[12pt]{article}

\usepackage{scicite}
\usepackage{graphicx}
\usepackage{times}
\usepackage{amsfonts}
\usepackage{amsmath}
\usepackage{float}
\usepackage{color, soul}
\usepackage{diagbox}
\usepackage{caption}
\usepackage{hyperref}
\usepackage{titling}
\usepackage{pdfpages}
\usepackage{amssymb} 
\usepackage{booktabs}
\usepackage{lineno}
\usepackage{authblk}
\usepackage{longtable}

\usepackage[superscript, biblabel, nomove]{cite}

\newenvironment{sciabstract}{%
\begin{quote} \bf}
{\end{quote}}

\title{Large language models surpass human experts in predicting neuroscience results}

\author[1, †, *]{\mbox{Xiaoliang Luo}}
\author[1, †]{\mbox{Akilles Rechardt}}
\author[2, †]{\mbox{Guangzhi Sun}}
\author[3, 6, †]{\mbox{Kevin K. Nejad}}
\author[4, †]{\mbox{Felipe Yáñez}}
\author[5, †]{\mbox{Bati Yilmaz}}
\author[8]{\mbox{Kangjoo Lee}}
\author[9]{\mbox{Alexandra O. Cohen}}
\author[10]{\mbox{Valentina Borghesani}}
\author[11, 12, 13]{\mbox{Anton Pashkov}}
\author[14]{\mbox{Daniele Marinazzo}}
\author[15]{\mbox{Jonathan Nicholas}}
\author[16]{\mbox{Alessandro Salatiello}}
\author[17]{\mbox{Ilia Sucholutsky}}
\author[18]{\mbox{Pasquale Minervini}}
\author[19]{\mbox{Sepehr Razavi}}
\author[20]{\mbox{Roberta Rocca}}
\author[21]{\mbox{Elkhan Yusifov}}
\author[22]{\mbox{Tereza Okalova}}
\author[23]{\mbox{Nianlong Gu}}
\author[24]{\mbox{Martin Ferianc}}
\author[25]{\mbox{Mikail Khona}}
\author[26, 27]{\mbox{Kaustubh R. Patil }}
\author[28, 29]{\mbox{Pui-Shee Lee}}
\author[30]{\mbox{Rui Mata}}
\author[31]{\mbox{Nicholas E. Myers}}
\author[32]{\mbox{Jennifer K Bizley}}
\author[33, 34]{\mbox{Sebastian Musslick}}
\author[35]{\mbox{Isil Poyraz Bilgin}}
\author[36]{\mbox{Guiomar Niso}}
\author[37]{\mbox{Justin M. Ales}}
\author[38]{\mbox{Michael Gaebler}}
\author[39]{\mbox{N Apurva Ratan Murty}}
\author[40]{\mbox{Leyla Loued-Khenissi}}
\author[41]{\mbox{Anna Behler}}
\author[42, 43]{\mbox{Chloe M. Hall}}
\author[44, 45]{\mbox{Jessica Dafflon}}
\author[46]{\mbox{Sherry Dongqi Bao}}
\author[1, 7, †]{\mbox{Bradley C. Love}}
\affil[ ]{%
$^{1}$Department of Experimental Psychology, University College London, London, United Kingdom, $^{2}$Department of Engineering, University of Cambridge, Cambridge, United Kingdom, $^{3}$Department of Physiology, Anatomy \& Genetics, University of Oxford, Oxford, United Kingdom, $^{4}$Max Planck Institute for Neurobiology of Behavior – caesar, Bonn, Germany, $^{5}$National Magnetic Resonance Research Center (UMRAM), Bilkent University, Ankara, Turkey, $^{6}$Department of Computer Science, University of Bristol, Bristol, United Kingdom, $^{7}$The Alan Turing Institute, London, United Kingdom, $^{8}$Department of Psychiatry, Yale University, New Haven, United States, $^{9}$Psychology, Emory University, Atlanta, United States, $^{10}$Faculty of Psychology and Educational Sciences, Université de Genève, Genève, Switzerland, $^{11}$Neurosurgery, Novosibirsk State Medical University, Novosibirsk, Russian Federation, $^{12}$Federal Center of Neurosurgery, FSBI, Novosibirsk, Russia, $^{13}$ Department of Data Collection and Processing Systems, Novosibirsk State Technical University, Novosibirsk, Russia, $^{14}$Department of Data Analysis, Ghent University, Ghent, Belgium, $^{15}$Psychology, New York University, New York, United States, $^{16}$Department of Cognitive Neurology, University of Tübingen, Tübingen, Germany, $^{17}$Computer Science, Princeton University, Princeton, United States, $^{18}$ILCC, University of Edinburgh, Edinburgh, United Kingdom, $^{19}$Philosophy, Psychology, and Language Sciences , The University of Edinburgh , Edinburgh , United Kingdom, $^{20}$Department of Culture, Cognition and Computation, Aarhus University, Aarhus, Denmark, $^{21}$Department of Molecular Life Sciences, University of Zurich, Zurich, Switzerland, $^{22}$Bioengineering, University of Pennsylvania, Philadelphia, United States, $^{23}$Linguistic Research Infrastructure, University of Zurich, Zurich, Switzerland, $^{24}$Electronic and Electrical Engineering, University College London, London, United Kingdom, $^{25}$Brain and Cognitive Sciences, Massachusetts Institute of Technology, Cambridge, Massachusetts, United States, $^{26}$Institute of Neuroscience and Medicine, INM-7: Brain and Behaviour, Research Centre Jülich, Jülich, Germany, $^{27}$Medical Faculty, Institute of Systems Neuroscience, Heinrich Heine University Düsseldorf, Düsseldorf, Germany, $^{28}$Graduate School of Systemic Neurosciences, Ludwig-Maximilians-University Munich, Planegg-Martinsried, Germany, $^{29}$Institute of Neuronal Cell Biology, Technical University of Munich, Munich, Germany, $^{30}$Faculty of Psychology, University of Basel, Basel, Switzerland, $^{31}$School of Psychology, University of Nottingham, Nottingham, United Kingdom, $^{32}$Ear Institute, University College London, London, United Kingdom, $^{33}$Institute of Cognitive Science, University of Osnabrück, Osnabrück, Germany, $^{34}$Department of Cognitive, Linguistic, \& Psychological Sciences, Brown University, Providence, United States, $^{35}$Département de psychologie, Centre de recherche de l'Institut universitaire de gériatrie de Montréal, Montreal, Canada, $^{36}$Instituto Cajal, CSIC, Madrid, Spain, $^{37}$School of Psychology \& Neuroscience, University of St Andrews, St Andrews, United Kingdom, $^{38}$Neurology, Max Planck Institute for Human Cognitive and Brain Sciences, Leipzig, Germany, $^{39}$Department of Psychology, Georgia Institute of Technology, Atlanta, United States, $^{40}$Département des Neurosciences Cliniques, Lausanne University Hospital, Lausanne, Switzerland, $^{41}$School of Psychological Science, The University of Newcastle, Newcastle, Australia, $^{42}$Institute of Physiology, University Medical Center of the Johannes Gutenberg University, Mainz, Germany, $^{43}$Institute for Quantitative and Computational Biosciences (IQCB), Johannes Gutenberg University, Mainz, Germany, $^{44}$Data Science and Sharing Team, Functional Magnetic Resonance Imaging Facility, National Institute of Mental Health, Bethesda, United States, $^{45}$Machine Learning Team, Functional Magnetic Resonance Imaging Facility, National Institute of Mental Health, Bethesda, United States, $^{46}$Zurich Center for Neuroeconomics, Department of Economics, University of Zurich, Zurich, Switzerland, †Major contributions, the names of the remaining authors are listed in random order, see contributions breakdown in Table \ref{table:contributions}, *Corresponding author: xiao.luo.17@ucl.ac.uk}

\date{}

\begin{document} 
\baselineskip24pt
\maketitle
\newpage
\begin{sciabstract}
Scientific discoveries often hinge on synthesizing decades of research, a task that potentially outstrips human information processing capacities. Large language models (LLMs) offer a solution. LLMs trained on the vast scientific literature could potentially integrate noisy yet interrelated findings to forecast novel results better than human experts. To evaluate this possibility, we created BrainBench, a forward-looking benchmark for predicting neuroscience results. We find that LLMs surpass experts in predicting experimental outcomes. BrainGPT, an LLM we tuned on the neuroscience literature, performed better yet. Like human experts, when LLMs were confident in their predictions, they were more likely to be correct, which presages a future where humans and LLMs team together to make discoveries. Our approach is not neuroscience-specific and is transferable to other knowledge-intensive endeavors.

\end{sciabstract}

\section*{Introduction}
Keeping up with the exponentially increasing \cite{bornmann_growth_2015} scientific literature is a superhuman challenge. Potentially disruptive findings go unnoticed in the deluge of articles \cite{chu_slowed_2021}. Processing and integrating the myriad of relevant findings may already surpass humans' abilities.
One path forward is a partnership between human scientists and machines. This partnership could take several forms, including specialist solutions that address specific challenges, such as in protein folding \cite{tunyasuvunakool_highly_2021}, drug discovery \cite{zhavoronkov_deep_2019}, and materials science \cite{tshitoyan_unsupervised_2019}. Alternatively, general models of the scientific literature could help guide human scientists' predictions and study designs. We consider this possibility.

It is an open question whether large language models (LLMs), trained on general text and scientific articles, can predict the outcomes of experiments. If LLMs' predictions surpassed human experts, the practice of science and the pace of discovery would radically change. We consider this question for neuroscience, which is a large and interdisciplinary field. Prediction in neuroscience should be challenging for human experts for several reasons: (i) there are often many thousands of relevant scientific articles, (ii) an individual study can be noisy, unreliable, and may not replicate, (iii) neuroscience is a multi-level endeavor \cite{mok_multilevel_2023}, spanning behavior and molecular mechanisms, (iv) and the analysis methods are diverse and can be complex \cite{botvinik-nezer_variability_2020}, (v) as are the methods used, which include different brain imaging techniques, lesion studies, gene modification, pharmacological interventions, and so forth.

Can LLMs meet these challenges? In other domains, LLMs have performed impressively. Upon its release, OpenAI's ChatGPT \cite{liu_summary_2023} captured the public's imagination with its abilities. Most LLMs are based on the transformer architecture \cite{vaswani_attention_2017}. These models
contain billions and sometimes trillions of weights \cite{fedus_switch_2021}, which are tuned during training in a self-supervised manner to predict the next token, such as the next word in a text passage.

LLMs have displayed remarkable capabilities, including passing professional exams, reasoning (though not without limitations), translation, solving mathematics problems, and even writing computer code \cite{srivastava_beyond_2022,gunasekar_textbooks_2023}. By constructing a statistical model during their training to predict the next token, whether that token is a word, pixel, or protein sequence \cite{strack_visual_2023}, LLMs uncover the underlying patterns or structure of a domain. This generative model captures patterns in the training data, including subtle and imperfect ones. How LLMs learn and generalize to novel situations can be likened to how expert scientists detect patterns in their field after years of reading papers, attending conferences, and analyzing data. From these experiences, human experts build intuitions that enable them to predict future outcomes based on proposed study designs. However, unlike human scientists, LLMs are virtually unconstrained in how much of the scientific literature they can process during training.

\begin{figure}
    \centering
    \includegraphics[width=0.7\textwidth]{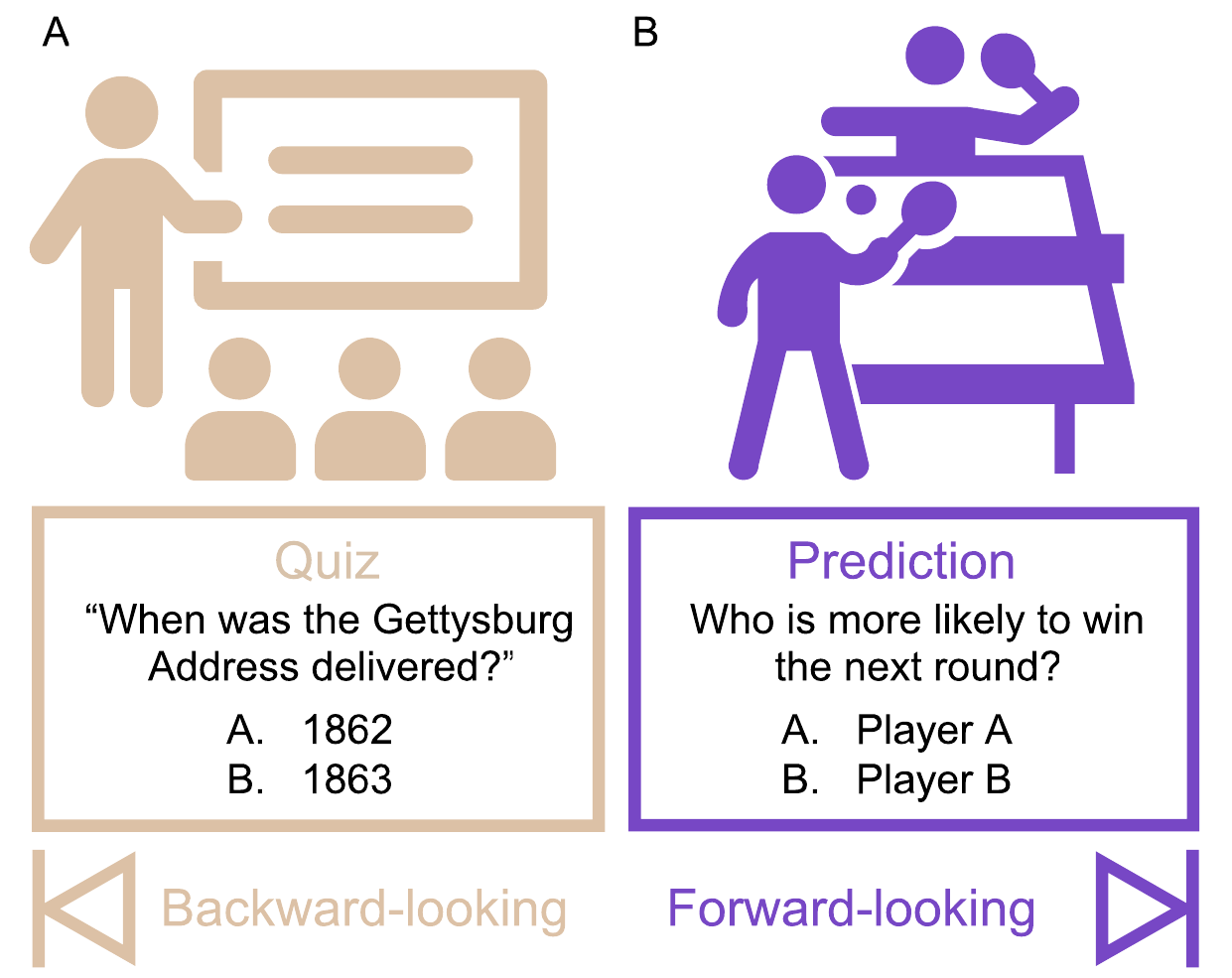}
    \caption{\textbf{Backward-looking and Forward-looking evaluations.} (A) Backward-looking benchmarks involve recalling factual information. For example, in the left panel, a student retrieves a fact about the Gettysburg Address that they learned during a history class. Existing benchmarks in scientific domains are in essence backward-looking as they emphasize retrieving accepted facts for question answering and reasoning tasks.
    (B) Forward-looking benchmarks involve predicting novel outcomes based on past data. Two forms of uncertainty, aleatoric (due to intrinsic randomness) and epistemic (due to lack of knowledge), may be present. For example, in the right panel, a table tennis fan predicts which player will win the next set based on their knowledge of the players, how they have played so far today, and so forth. Inherent random factors, such as a breeze affecting the ball's flight, will also be present.}
    \label{fig:future_looking}
\end{figure}

How can we formally evaluate the predictive abilities of LLMs in neuroscience? With the rise of LLMs, there has been a surge in evaluation benchmarks, many of which focus on assessing LLMs' capabilities in scientific domains. Most benchmarks evaluate core knowledge retrieval and reasoning abilities, which are typically \textit{backward-looking} (Fig. \ref{fig:future_looking}). Backward-looking benchmarks include MMLU \cite{hendrycks_measuring_2021}, PubMedQA \cite{jin_pubmedqa_2019}, and MedMCQA \cite{pal_medmcqa_2022}. These benchmarks are structured in a question-and-answer format, where models must demonstrate extensive world knowledge, retrieve relevant information based on the context of the question, and answer correctly. However, none of these benchmarks are suitable for evaluating the ability of models to predict novel outcomes, which is inherently \textit{forward-looking} (Fig. \ref{fig:future_looking}).

\begin{figure}
    \centering
    \includegraphics[width=\textwidth]{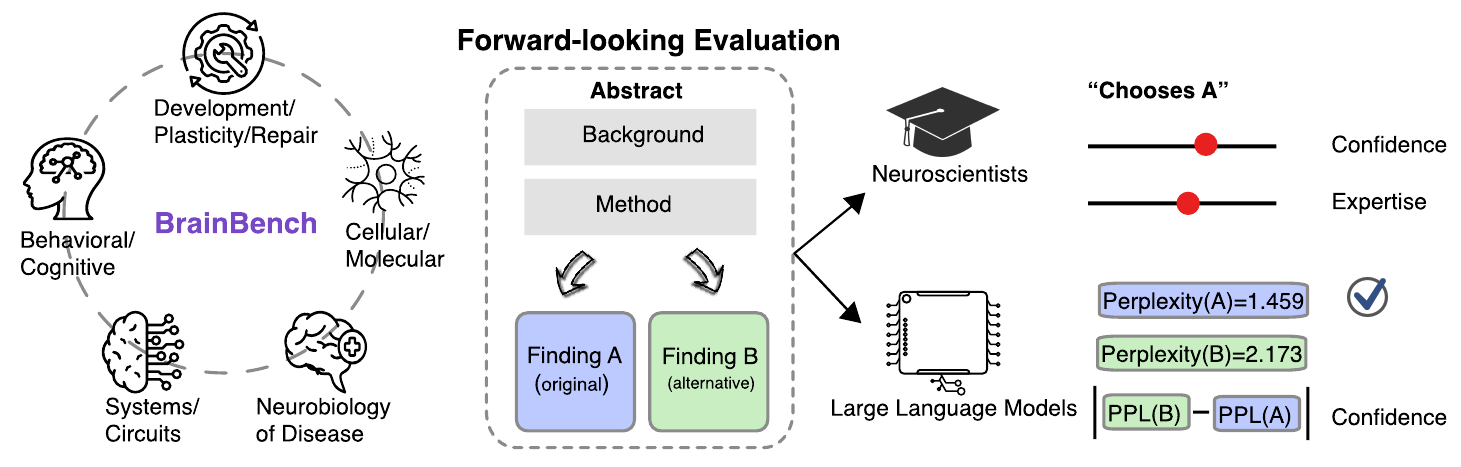}
    \caption{\textbf{BrainBench is a forward-looking benchmark for neuroscience.} BrainBench evaluates test-takers' ability to predict neuroscience results.
    BrainBench's test cases were sourced from recent {\em Journal of Neuroscience} abstracts across five neuroscience domains: Behavioral/Cognitive, Systems/Circuits, Neurobiology of Disease, Cellular/Molecular, and Developmental/Plasticity/Repair. Test-takers chose between the original abstract and one altered to significantly change the result while maintaining coherency. Human experts and Language Models (LLMs) were tasked with selecting the correct (i.e., original) version from the two options. Human experts made choices, and provided confidence and expertise ratings in an online study. LLMs were scored as choosing the abstract with the lower perplexity (i.e., the text passage that was less surprising to the model) and their confidence was proportional to the difference in perplexity between the two options.}
    \label{fig:brainbench}
\end{figure}

To address this need, we developed BrainBench to test LLMs' ability to predict neuroscience findings (Fig. \ref{fig:brainbench}). LLMs have been trained extensively on the scientific literature, including neuroscience. BrainBench evaluates whether LLMs have seized on the fundamental patterning of methods and results that underlie the structure of neuroscience. Can LLMs outperform human experts on this forward-looking benchmark? In particular, BrainBench evaluates how well the test-taker can predict neuroscience results from methods by presenting two versions of an abstract from a recent journal article. The test-taker's task is to predict the study's outcome, choosing between the original and an altered version. The altered abstract significantly changes the study's outcome (i.e., results) while maintaining overall coherence. 

To appreciate how BrainBench qualitatively differs from existing benchmarks, consider a perceived limitation of LLMs, namely their tendency to generate erroneous information, a phenomenon commonly referred to as ``hallucination" by LLM researchers. Unlike knowledge graphs that store verified facts, LLMs may not be trustworthy for backward-looking tasks such as summarizing research papers or providing accurate citations \cite{lewis_retrieval-augmented_2021}. However, for forward-looking tasks, such as predicting results from a novel experiment, we view this tendency to mix and integrate information from large and noisy datasets as a virtue. What is a hallucination in a backward-looking task is a generalization or prediction in a forward-looking task (e.g., BrainBench). BrainBench provides a way to quantify this forward-looking ability and compare to human experts. To foreshadow our results, LLMs surpassed human experts on BrainBench by a substantial margin and this margin increased when we provided additional training in neuroscience to an LLM, which we refer to as ``BrainGPT".

\section*{Results}
\subsection*{General-purpose LLMs best neuroscientists on BrainBench}

On each benchmark trial (see Fig. \ref{fig:brainbench}), both the LLMs and human experts were tasked with selecting which of two versions of an abstract was correct (i.e., the original version). Human neuroscience experts were screened for 
their expertise and engagement (see Methods) with $171$ out of $202$ participants passing all checks and included in our analyses.

\begin{figure}
    \centering
    \includegraphics[width=\textwidth]{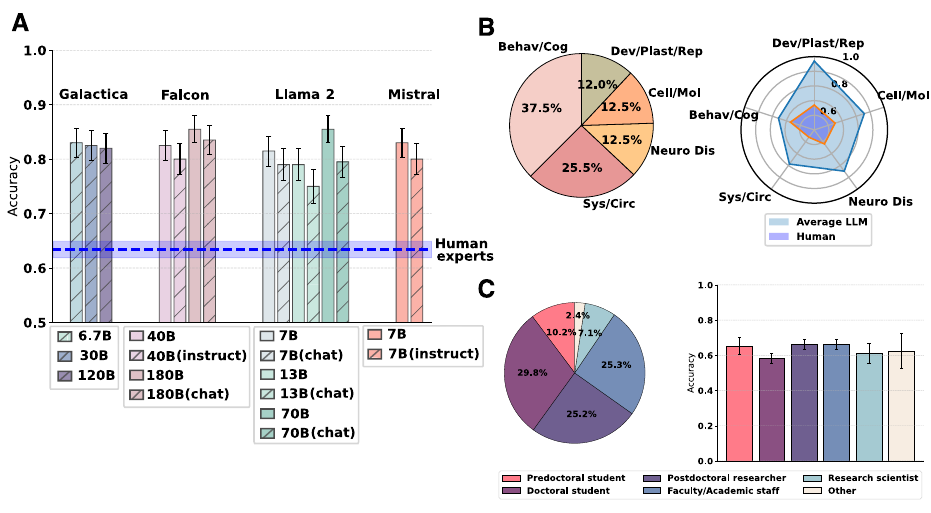}
    \caption{\textbf{Performance of human experts and large language models on BrainBench.} (A) LLMs outperformed human experts on BrainBench. Smaller models are on par with larger models. 
    Base versions of models outperformed chat and instruct versions, which were tuned to be conversational with humans.
    (B) The distribution of test cases across neuroscience subfields roughly mirrors the distribution of articles in the {\em Journal of Neuroscience} with Behavior/Cognitive overrepresented. The average performance of $15$ LLMs and human experts is shown. LLMs outperformed human experts in every subfield (see Fig. \ref{fig:breakdown_subfield} for full results). Error bars represent the standard deviation of accuracy around the mean. (C) Majority of the participants were doctoral students, postdoctoral researchers and faculty/academic staff. Error bars represent standard error of the mean.}
    \label{fig:results_3_in_1}
\end{figure}

Every LLM outperformed human experts on BrainBench with LLMs averaging $81.4\%$ accuracy and human experts averaging $63.4\%$ (Fig. \ref{fig:results_3_in_1}A). When restricting human responses to those in the top $20\%$ of self-reported expertise for that test item, accuracy rose to $66.2\%$, still below the level of LLMs.

Smaller models such as Llama2-7B and Mistral-7B with $7$ billion parameters, performed comparably to larger models (Fig. \mbox{\ref{fig:results_3_in_1}}A) while besting smaller models (see Supplementary Materials) that may lack the capacity to capture key data patterns. Chat or instruction-optimized models performed worse than their base model counterparts ($t(5)=5.4$, $p = .002$). We suspect that aligning LLMs to engage in natural language conversations hinders their scientific inference abilities (see Discussion).

The previous analyses involved benchmark items created by co-authors who are neuroscience experts (see Methods). We conducted the same analyses using test cases generated by a LLM, namely GPT-4 (see Methods), and observed similar results (see Supplementary materials). 

\paragraph{Performance breakdown by subfields and by participant types}
BrainBench encompasses test cases from five distinct neuroscience domains: Behavioral/Cognitive, Cellular/Molecular, Systems/Circuits, Neurobiology of Disease, and Development/Plasticity/Repair. Some domains, particularly Behavioral/Cognitive, are overrepresented both in BrainBench (Fig. \ref{fig:results_3_in_1}B) and the {\em Journal of Neuroscience} from which we drew our test cases (see Methods).

On average, LLMs performed better than human experts in every subfield  (Fig. \ref{fig:results_3_in_1}B), as did each individual LLM (Fig. \ref{fig:breakdown_subfield}). Most human experts were doctoral students, postdoctoral researchers, or faculty/academic staff (see Fig. \ref{fig:results_3_in_1}C). 
Please refer to Supplementary materials for more detailed demographic information including years of experience in neuroscience research about the human experts and distributions of self-reported expertise by subfields (Fig. \ref{fig:expertise_distr_per_subfield}).

\paragraph{Do judgments from LLMs and human experts align?}
We considered whether human experts and LLMs found the same benchmark items difficult. For humans, we calculated the mean accuracy for each of the 200 test cases. For LLMs, we calculated the signed differences in perplexity between incorrect and correct abstracts for each test case. 
Perplexity measures how surprising a text passage is to an LLM.
Using these measures (Fig. \ref{fig:error_correlation_heatmap}), the mean Spearman correlation between an LLM and human experts was $0.15$ ($\pm0.03$) whereas the mean Spearman correlation between LLMs was $0.75$ ($\pm0.08$).

\paragraph{LLMs can integrate information across context}
To better understand the basis for the remarkable performance of LLMs (see Fig. \ref{fig:iso} for results), we investigated whether their performance was achieved by integrating information throughout the abstract (including the method used) or by solely relying on the local context in the results passages that differed between the original and altered abstract (Fig. \ref{fig:brainbench})

We re-evaluated the LLMs on individual sentences containing only the altered results passage (i.e., local context only). LLMs performed much worse when restricted to this local context (Fig. \ref{fig:iso}), which provides strong evidence that LLMs are integrating information across the abstract, including information on background and methods. LLM's superior performance relative to human experts appears to arise from integrating information across the abstract.

In addition, we analyzed whether LLMs benefited from a general neuroscience context (similar to few-shot prompting) rather than integrating study-relevant information. We tested models using abstracts with sentences randomly swapped from within the same neuroscience subfield. Both original and altered abstracts were used to re-evaluate LLMs' performance. As shown in Fig. \mbox{\ref{fig:swap}}, there was a significant performance decline with coherent versus swapped contexts, indicating that LLMs only partially benefit from accurate, domain-specific, but non-study-relevant context.

\paragraph{LLM performance is not driven by data memorization}
When LLMs perform well on a benchmark, one general concern is that the benchmark itself was part of the training set, allowing the LLM to memorize the correct answers. To address this concern, we used a commonly applied measure, zlib-perplexity ratio, for evaluating whether LLMs have memorized passages \cite{carlini_extracting_2021}. This ratio gauges the difference between a data-agnostic compression rate of text and data-specific perplexity computed by an LLM (see Methods). Passages that are hard to compress but have low perplexity are indicative of memorization.

We found no indication that BrainBench was memorized by LLMs (Fig. \ref{fig:memorisation}). For comparison, we calculated the zlib-perplexity ratio for a passage that we suspected would be memorized by LLMs, namely the Gettysburg Address. The Gettysburg Address should appear multiple times in an LLM's training set and indeed it showed signs of memorization (Fig. \ref{fig:memorisation}). Interestingly, for some LLMs, we know exactly what they were trained on (see Table \ref{tab:data_sources}). For these models, the distribution of zlib-perplexity ratios heavily overlapped for items that we knew were in the training set and for items, including BrainBench, that we knew were not in the training set. We suspect that the overlap may indicate scientific articles, which are unlikely to repeat in training sets, are stored in LLMs as general patterns, similar to human schemas, supporting performance on tasks requiring generalization (e.g., BrainBench). This hypothesis invites future study.


As a final check (Fig. \ref{fig:pubdates_difficulty_correlation}; Materials and Methods), we confirmed that LLMs do not perform better on items published earlier in 2023 (e.g., Jan-2023 vs. Oct-2023), which addresses the concern that early items are more likely to have a preprint or other precursor appear in the training set that affected BrainBench performance. All our checks indicated that BrainBench items were novel for the LLMs.

\subsection*{LLMs and human experts are calibrated}
\begin{figure}
    \centering
    \includegraphics[width=.8\textwidth]{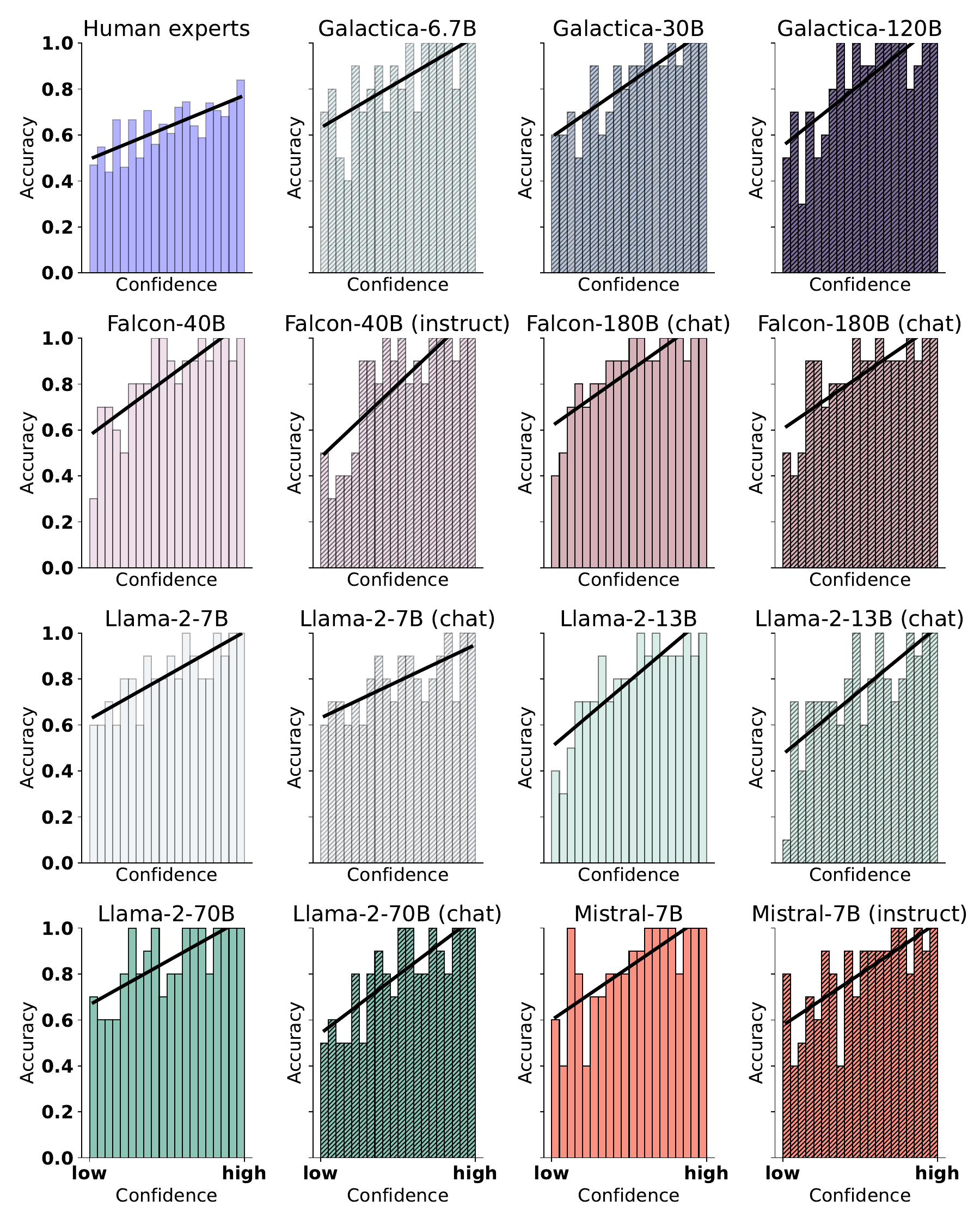}
    \caption{\textbf{Accuracy and confidence are calibrated for human experts and large language models (LLMs).} When human experts and LLMs are confident in their BrainBench judgments, they are more likely to be correct. Confidence ratings were sorted and placed in equally-sized bins with the mean accuracy for items in that bin plotted. The positive slope of the black regression lines for human experts and all LLMs indicates that confidence is well calibrated (i.e., higher confidence corresponds to higher accuracy).
    Calibration is beneficial for human-machine teams.}
    \label{fig:calibration}
\end{figure}

To assess whether LLMs' predictions are calibrated, we examined how well their confidence tracked their accuracy, a crucial characteristic for a trustworthy prediction system. We estimated LLMs' confidence using the ranked absolute difference in perplexities between two abstracts (Fig. \ref{fig:brainbench}; see Methods) and found, like human experts, all LLMs exhibited a positive correlation between accuracy and confidence. When LLMs are confident in their decisions, they are more likely to be correct (Fig. \ref{fig:calibration}). In addition, we fitted logistic regressions between model perplexity differences to their correctness as well as human confidences to their correctness on the individual level. We observed significant positive correlations, confirming both models and humans are calibrated (Table \ref{tab:lr_calibration}).

\subsection*{Augmenting LLMs with neuroscience knowledge to create BrainGPT}
Pretrained LLMs can provide a foundation for further training in neuroscience with the aim of improving performance, as assessed by BrainBench. We used Low-Rank Adaptation (LoRA) \cite{hu_lora_2021} to augment a pretrained LLM, Mistral-7B-v0.1, with additional neuroscience knowledge.

\begin{figure}
    \centering
    \includegraphics[width=.8\textwidth]{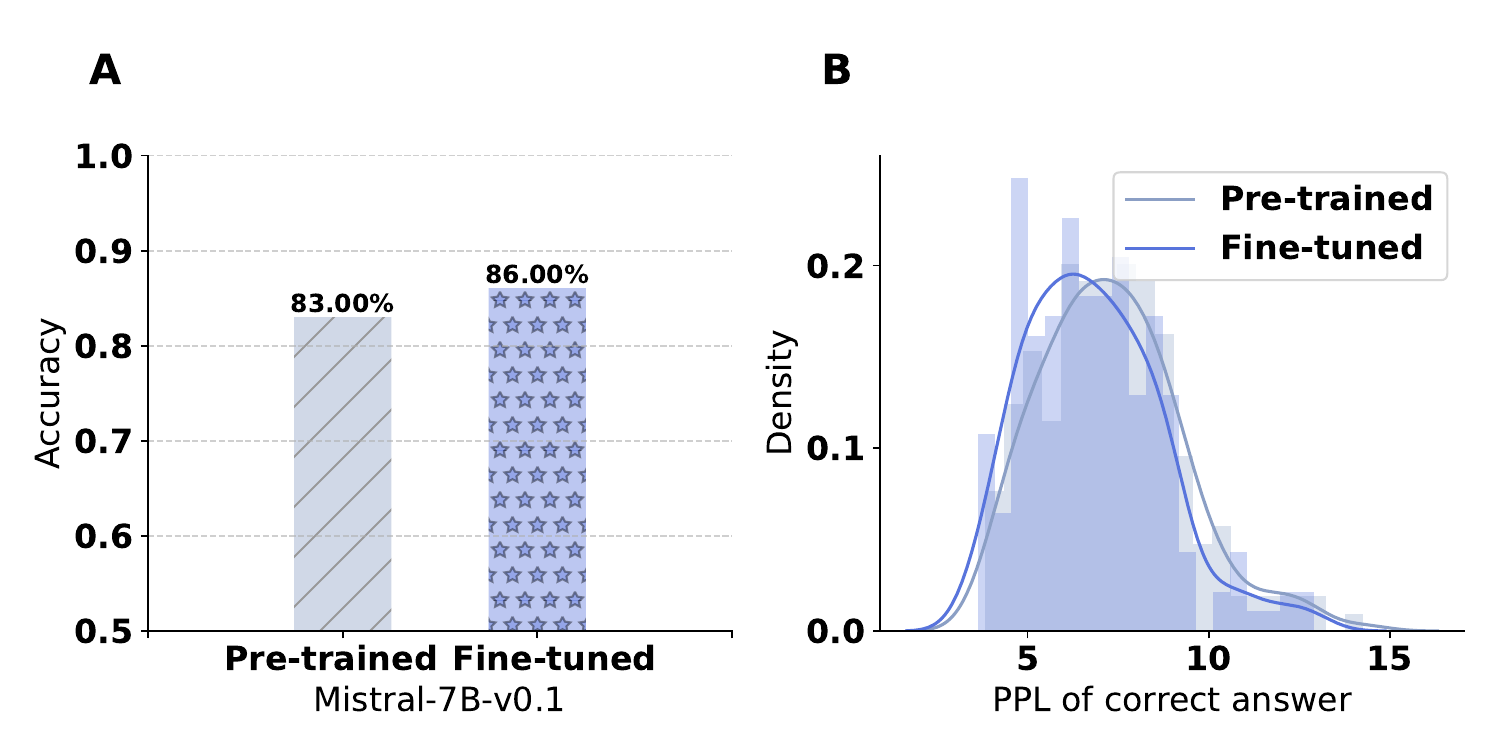}
    \caption{\textbf{Fine-tuning a pretrained large language model (LLM) on neuroscience knowledge.} Mistral-7B-v0.1 was fine-tuned using LoRA on neuroscience articles from 2002-2022 (a total of $1.3$ billion tokens). (A) The fine-tuned model improved by $3\%$ on BrainBench.
 (B) The fine-tuning process substantially shifted the perplexity distribution of correct responses, indicative of the LLM specializing in neuroscience.}
    \label{fig:finetune_boost}
\end{figure}

LoRA is a parameter-efficient fine-tuning technique that inserts low-rank adapter matrices into LLM transformer blocks (Fig. \ref{fig:lora}) and trains only these LoRA weights to update the model's behavior. In our case, we fine-tuned Mistral-7B-v0.1 using over $1.3$ billion tokens from neuroscience publications spanning $100$ journals between 2002 and 2022 (see Methods), which significantly improved performance by 3\% on BrainBench (Fig. \mbox{\ref{fig:finetune_boost}}A).

LoRA tuning dramatically shifted ($t(199)=15.7, p<.001$) the perplexity of correct responses (Fig. \ref{fig:finetune_boost}B), which is indicative of the LLM specializing for neuroscience material. LoRA introduced $629,145,600$ new weights, which is 8\% of the total number of weights in Mistral-7B-v0.1. These results indicate that BrainGPT models can efficiently be derived by extending existing LLMs.

\section*{Discussion}
We considered whether large language models (LLMs) can forecast the outcome of neuroscience experiments. By training on the vast scientific literature, we hoped LLMs could build a generative model that captured the patterns underlying neuroscience. To evaluate this possibility, we constructed a new forward-looking (Fig. \ref{fig:brainbench}) benchmark, BrainBench.

BrainBench assesses a test taker's ability to select which of two versions of a neuroscience abstract contains the actual results of the study (see Fig. \ref{fig:brainbench}). We found that LLMs outperform human experts on BrainBench by a considerable margin (see Fig. \ref{fig:results_3_in_1}A) across all neuroscience subfields we consider (Fig. \ref{fig:results_3_in_1}B). Moreover, the LLMs knew when their predictions were likely to be right or wrong (Fig. \ref{fig:calibration}). LLMs' superior performance arose from their ability to integrate information throughout the abstract, such as text pertaining to the method and study design. When access to such information was removed, LLM performance drastically declined (Fig. \ref{fig:iso}).

We found no indication that LLMs had been exposed to and memorized BrainBench items during their training. Instead, our analyses suggested that LLMs discovered the fundamental patterns that underlie neuroscience studies, which enabled LLMs to predict the outcomes of studies that were novel to them. These conclusions were supported by a widely employed technique \cite{carlini_extracting_2021} to determine text membership within an LLMs' training set (see Fig. \ref{fig:memorisation}). 
The  Galactica \cite{taylor_galactica_2022} LLMs were particularly illuminating because we know which articles were not in the training set vs. ones that might be. Interestingly, there was no indication of memorization in models such as Galactica for scientific articles that were in its training set, consistent with the notion that LLMs learn broad patterns underlying scientific fields. While passages that frequently repeat in the training set, such as the Gettysburg Address, may be memorized (see Fig. \ref{fig:memorisation}), scientific articles that occur infrequently (most likely once) in the training set appear to support LLM's forward-looking predictive abilities. As a final check, we trained a relatively small LLM from scratch on the published neuroscience literature (excluding preprints and BrainBench items), which eliminated any possible overlap between training data and BrainBench, and found superhuman performance on BrainBench (Fig. \mbox{\ref{fig:more}}).

LLM's impressive forward-looking capabilities suggest a future in which LLMs help scientists make discoveries.
To be effective, LLMs need to stay abreast of the rapidly expanding literature. 
 We found that LLMs could efficiently be augmented with neuroscience knowledge using LoRA \cite{hu_lora_2021}, boosting performance on BrainBench (Fig. \ref{fig:finetune_boost}). LoRA provides a way to create BrainGPT models by re-orienting general-purpose LLMs for use in neuroscience. One can easily imagine a future in which BrainGPT is near continuously updated with new knowledge using LoRA, along with complementary approaches such as Retrieval Augmented Generation (RAG; \cite{lewis_retrieval-augmented_2021}). RAG could be used to query a database of relevant and up to date scientific articles for the task at hand.

In addition to keeping LLMs up to date, benchmarks should routinely be refreshed and expanded to address current needs. One challenge is that creating forward-looking benchmarks, such as BrainBench, is labor intensive and requires human expertise. To address this potential bottleneck, we created and evaluated 100 test cases using GPT-4 through a largely automated process (see Methods). Although there is room for improvement, these items were close in quality to the human-created ones with 8 of the 100 items being word-for-word matches with the human-created versions. These efforts should pave the way for the rapid creation of other forward-looking benchmarks in neuroscience, as well as benchmarks for other knowledge intensive fields. We believe high-quality forward-looking benchmarks will be critical to developing LLMs as tools for scientific discovery.

For LLMs to be trustworthy and effective teammates, they need to convey the certainty of their predictions to human scientists. Fortunately, we found that LLMs' confidence is well calibrated. When LLMs were confident in their predictions, they were more likely to be correct (Fig. \ref{fig:calibration}). A second ingredient for effective teams is being diverse or complementary. LLMs have potential here as well as the items they found difficult did not highly correlate with those human experts found difficult (Fig. \ref{fig:error_correlation_heatmap}). These two ingredients, being well calibrated and complementary, allow systems that combine human and machine judgments to outperform either alone \cite{steyvers_bayesian_2022}.

All our results, including those for calibrated confidence, were only possible because we had access to LLM weights to calculate the perplexity of passages (see Fig. \ref{fig:brainbench}). Our approach diverged from the popular approach of prompting models for responses through natural language (i.e., chat). Prompting in natural language may yield less reliable judgments and degrade model competency compared to using model probability scores or training separate classifiers directly from internal representations \cite{zheng_judging_2023,hu_prompting_2023,azaria_internal_2023,gao_framework_2023}. These observations underscore the importance of working with models that are as open as possible, ideally making both the weights and training set publicly available. Accordingly, we make BrainGPT available on the Huggingface platform \href{https://huggingface.co/BrainGPT}{https://huggingface.co/BrainGPT}.

Beyond serving as a tool for neuroscientists, BrainGPT can help reveal the structure of the field. In particular, we can vary BrainGPT's training set and observe the effect on BrainBench. For example, what is the effect of including training data from related fields like psychology? In terms of supporting prediction, we can quantify how interrelated fields are. Does it help to weight articles in the training set by their recency, citations, or impact factor? In addition to these training manipulations, we can vary how testing is conducted. For example, would step-by-step thinking via chain-of-thought reasoning \cite{wei_chain--thought_2023} benefit BrainGPT? If prediction in neuroscience is akin to a deductive reasoning process, then it should. If instead, as we suspect, prediction in neuroscience is a function of many noisy intertwined signals across subfields, then chain-of-thought reasoning will not help. BrainGPT and BrainBench can help answer these meta-science questions. 

We foresee a future in which LLMs serve as forward-looking generative models of the scientific literature. LLMs can be part of larger systems that assist researchers in determining the best experiment to conduct next. One key step toward achieving this vision is demonstrating that LLMs can identify likely results. For this reason, BrainBench involved a binary-choice between two possible results. LLMs excelled at this task, which brings us closer to systems that are practically useful. In the future, rather than simply selecting the most likely result for a study, LLMs can generate a set of possible results and judge how likely each is. Scientists may interactively use these future systems to guide the design of their experiments.

One risk is that scientists do not pursue studies when their predictions run counter to those of an LLM. In some cases, this might be a sensible course of action, whereas in other cases the LLM might have identified potential gaps or errors in the scientific literature. In the latter situation, conducting the study might result in a significant breakthrough. Conversely, a study result that was predicted with high confidence by an LLM might be viewed as an incremental advance.

LLMs' predictions are informed by a vast scientific literature that no human could read in their lifetime. As LLMs improve, so should their ability to provide accurate predictions. In this contribution, we focused on neuroscience but our aims are broader – we hope to provide a template for any knowledge intensive field. None of the approaches we adopted are neuroscience-specific. Indeed, the degree of efficacy of our approach may depend on the underlying structure of the domain. For instance, disciplines like mathematics, which rely heavily on logical deduction, might not benefit as much as other scientific fields that involve pattern-based reasoning.

We hope to democratize the use of LLMs in science and increase reproducibility by highlighting the use of relatively small models that can be run locally and whose weights are accessible, which contrasts with commercial products. Finally, while LLMs appear poised to supplant humans at prediction, we foresee a role for human experts in providing the accompanying scientific explanations. Prediction is very important, but not everything.

\section*{Data availability}
Human participant data, and intermediate data generated via simulations and analyses are publicly available at \href{https://github.com/braingpt-lovelab/BrainBench}{https://github.com/braingpt-lovelab/BrainBench}. Model weights and training data are available at \href{https://huggingface.co/BrainGPT}{https://huggingface.co/BrainGPT}.

\section*{Code availability}
All computer code associated with this work including model training, evaluation, data processing and analyses are publicly available at \href{https://github.com/braingpt-lovelab/BrainBench}{https://github.com/braingpt-lovelab/BrainBench}.

\section*{Acknowledgments}
This work was supported the ESRC (ES/W007347/1), Microsoft (Accelerate Foundation Models Research Program), and a Royal Society Wolfson Fellowship (18302) to B.C.L. We thank Mona Garvert, Pradeep Reddy Raamana, Todd Hare, Yoav Kessler, Oliver Robinson and D.R. for their assistance. We thank the participants of the online study: Gaia Molinaro, Judy Zhu, Majd Abdallah, Yuri G. Pavlov, Jong-eun Lee, Adam Harris, Zhaoning Li, Roman Kessler, Lexi Zhang, Maciej Szul, Pranjul Gupta, Sunreeta Bhattacharya, Jellina Prinsen, Celine Gallagher, Michael Anes, Maarten Laroy, Tobias Ackels, Carina Forster, Pedro Gonçalves, Tommy Mcconnell, Diane Whitmer, Debottam Kundu, Benjamin Pasquereau, Jeremy Manning, Maciej Szul, Ahmed Hussain, Nicolas Clairis, Ignacio Vega-Vásquez, Kun Chen, Jeremy Hogeveen, Sina  Salehi, Suseendrakumar  Duraivel , Edgar Guevara, Ziyao Zhang, Thomas J. Younts, Marek Muszyński, Leonardo Dalla Porta, Todd Gureckis, Parnian Rafei, Feng-Chun Chou, Keith Temple, Alp Altunkaya, Andrew Tan, Jin Ho Yun, Arnau Marin-Llobet, Brian Lord, Daniel Lindh, Simon Besson-Girard, Eren Irmak, Emin Çelik, Aman Maharjan, Irene Sophia Plank. 

\section*{Declaration of interests}
The authors declare no competing interests.

\section*{Author contributions}
XL and BCL were responsible for primary writing. Test case creation was handled by BY, IPB, AP, TO, AOC, FY, EY, ARM, KL, VB, SR, JMA, RM, MG, GN, LLK, AB, KRP, MK, RR, KKN, AS, JN, DM, CMH, PSL, SM, NEM, JKB, SDB, NG, JD, and BCL. Quality control was conducted by BY, KKN, DM, PSL, NG, CMH, KL, SM, AR, ARM, IS, GN, IPB, RM, TO, MK, JMA, MG, EY, LLK, JD, JN, FY, RR, VB, SDB, AOC, AS, XL, and BCL. GPT-4 case creation was managed by KKN, XL, and BCL. Human-machine teaming involved FY, XL, and BCL. LoRA fine-tuning was performed by GS, XL, and BCL. Model evaluation was executed by XL, GS, MF, and BCL. Building the experiment was carried out by AR, XL, and BCL. Data analysis was done by XL, AR, and BCL. Conceptualization and strategy were undertaken by XL and BCL. Figure creation was the work of XL, BY, IB, CMH, GN, and BCL. Useful input and suggestions on the project were provided by all authors. Commenting and editing on the manuscript were also done by all authors. For a table breakdown, see Table \ref{table:contributions}.

\newpage 
\section*{Materials and Methods}
\renewcommand\thefigure{S.\arabic{figure}}  
\setcounter{figure}{0}  
\renewcommand{\thetable}{S.\arabic{table}}
\setcounter{table}{0}

We confirm that our research complies with all relevant ethical regulations; Experimental Psychology Ethics Board (University College London) approved the study protocol (ethics protocol: EP/2017/011); We confirm that informed consent was obtained from all human participants; Participant compensation is not applicable to the current study; None of our studies were preregistered.

\subsection*{Dataset Creation}
\begin{figure}[H]
    \centering
    \includegraphics[height=\textheight,width=\textwidth,keepaspectratio]{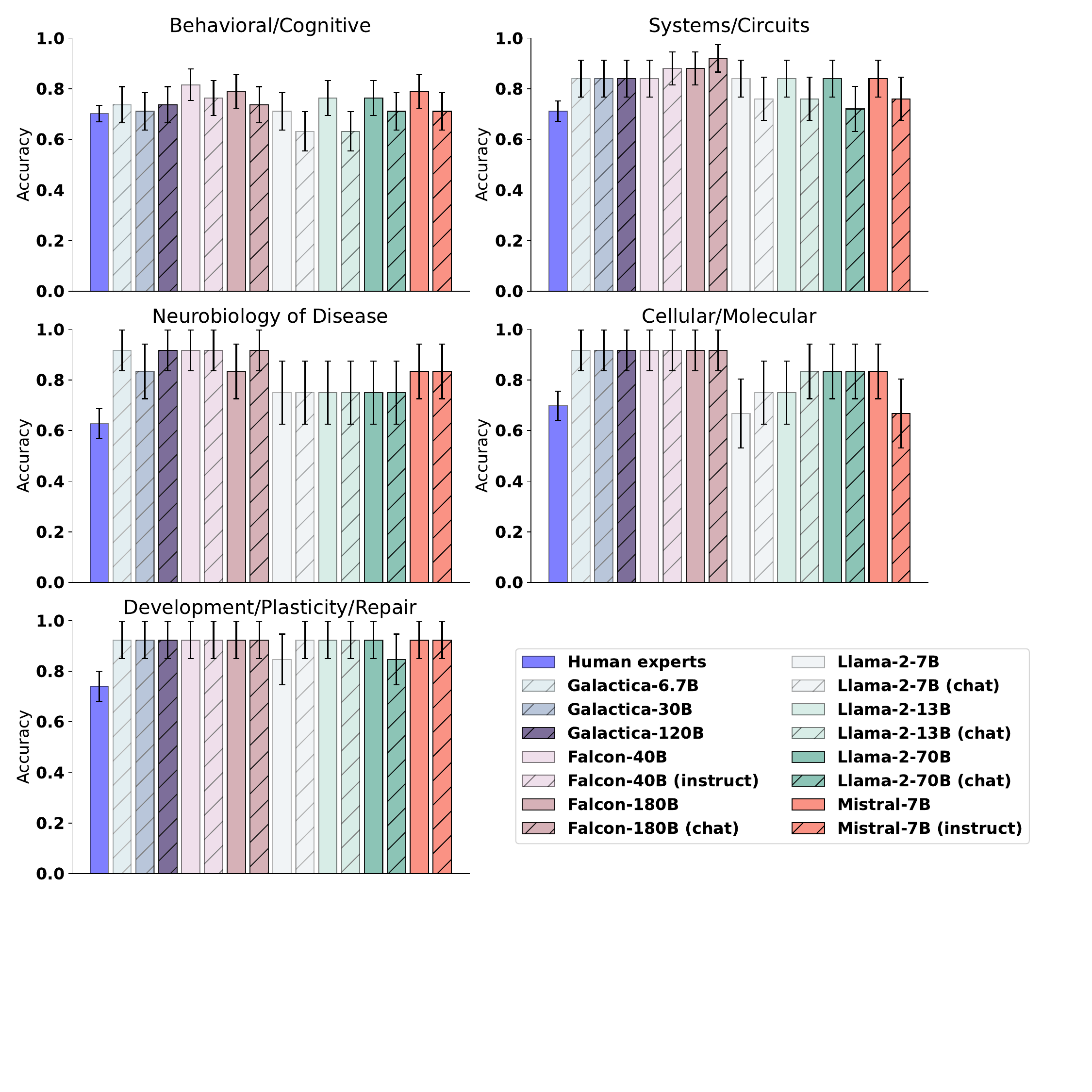}
    \caption{\textbf{BrainBench performance breakdown (using GPT-4 created test cases).} Accuracy by subfields of neuroscience across human experts and LLMs.}
    \label{fig:breakdown_subfield_machineCreated}
\end{figure}

Co-authors (Table \ref{table:contributions}) and GPT-4 (Azure OpenAI API; version 2023-05-15) created test cases that formed BrainBench. 
All test cases were sourced from \textit{Journal of Neuroscience} abstracts published in 2023 under the Creative Commons Attribution 4.0 International License (CC-BY). The abstracts are organized into five sections, namely Behavioral/Cognitive, Systems/Circuits, Neurobiology of Disease, Development/Plasticity/Repair, and Cellular/Molecular. In constructing BrainBench, we incorporated a total of 200 test cases crafted by human experts and an additional 100 test cases generated by GPT-4 (Azure OpenAI API; version 2023-05-15). All test cases were subjected to extensive quality control by human experts and GPT-4. For the distribution of test cases among subfields, refer to Fig. \ref{fig:results_3_in_1} for human-created cases and Fig. \ref{fig:breakdown_subfield_machineCreated} for GPT-4 generated cases. 

To create a test case, a published abstract was modified to create an altered version. The altered version significantly changed the results without changing the methods and background. Minimal changes were made that changed the basic result. For example, the altered abstract, compared to the original, could switch around the role of two brain regions in the results, reverse the direction of a result (e.g., replace “decreases” with “increases”), etc. Any changes maintained the coherency of the abstract which sometimes required multiple changes (e.g., replacing multiple “decreases” with “increases”). In other words, the altered abstracts needed to be empirically different, but not logically incoherent. Both volunteers and GPT-4 are given instructions that follow the essential criteria above. The exact wordings to prompt GPT-4 were slightly adjusted in order to obtain good quality test cases. We include the instructions used GPT-4 verbatim below. 

\paragraph{GPT-4 test creation prompt}
``Your task is to modify an abstract from a neuroscience research paper such that the changes significantly alter the result of the study without changing the methods and background. This way we can test the Artificial Intelligence understanding of the abstract’s subject area. 

Please read the instructions below and ensure you follow them one by one while you are modifying the abstracts:

- The format to submit is putting double brackets around the change with the first element being the original and the second element being your edit. E.g., [[original passage, modified passage]]. Always remember to wrap your edits with the double brackets; there should not be any other edits outside the brackets to the original abstract. 
- If you change a single word, never wrap the entire sentence inside the double brackets. For example, “... exhibit [[enhanced LTP and deficits in LTD, impaired LTP and enhanced LTD]].” is a wrong format, the correct format is: “... exhibit [[enhanced , impaired ]] LTP and [[deficits, enhanced]] in LTD.”
- The beginning of an abstract is the background and methods, so you should not alter those parts of the abstract. Do not alter the first couple sentences.
- We want the abstract to become empirically wrong, but not logically incoherent.
- To find the original result of the paper, one should require some neuroscience insight, not just general reasoning ability. So it is critical that the changes you make don’t evaluate the Artificial Intelligence reasoning ability, but its knowledge of neuroscience and how the brain works.
- Watch out for making changes that alter the results, but may still have occurred in the authors’ study. For example, an fMRI abstract on learning might mention the hippocampus and not the striatum. Nevertheless, the striatum might have also been active and not reported in the abstract because it was not the focus of the study. 
- The changes you make should not be identifiable or decodable from the rest of the abstract. Hence, if you make a change, make sure you change everything that can reveal the original abstract. For example, “activation of neurons in the visual cortex [[increases, decreases]] the activity in the motor cortex. This decrease in the activity of the visual cortex was followed by an increase in task performance.”. In this case it is very clear that the correct word is “decreases” as the next sentence (“This decrease in the activity of the visual cortex”) reveals that.
- Be mindful of the article when you change words. For example, if you change the word “decline” to “enhancement”, you must change the article as well, so the change will be [[a decline, an enhancement]].
- Ensure that your edits maintain inter-sentence consistency and proper syntax. The changes should not contradict or confuse the overall meaning of the abstract. 
- Avoid making trivial edits that do not require understanding of scientific concepts. The edits should reflect a deep understanding of the subject matter.
- Do not miss any crucial results or findings in the abstract while making the edits. Every significant point should be addressed in your modifications.

To generate better responses, you can use the topic of their study and purpose of studies in those topics. This knowledge helps you to find what modification you should do in the abstract.
Topics are:
- Behavioral/Cognitive: To understand how the brain influences behavior, cognition, and emotion, and to apply this understanding in diagnosing and treating neurological and psychiatric disorders.

- Cellular/Molecular: To study are to understand the functions and mechanisms of neurons at a cellular and molecular level, which includes investigating the biology of nerve cells, their genetic makeup, and how they form complex circuits, ultimately contributing to our understanding of brain function, behavior, and the development of treatments for neurological disorders.

- Neurobiology of Disease: To understand the biological basis of various neurological and psychiatric disorders in order to develop effective treatments and preventative measures.

- Development/Plasticity/Repair: to understand the mechanisms of brain development, adaptation, and repair in response to injury or disease, with the goal of developing strategies and treatments to enhance brain recovery and function.

- Systems/Circuits: to understand how neural circuits in the brain interact and coordinate to process information, control behavior, and support cognitive functions.

Here are two examples of the edited abstract by human experts which can help you to understand the task:

Example 1:
\verb|<|example\_1\verb|>|

Example 2:
\verb|<|example\_2\verb|>|

These are some common mistakes you have made in the past. So keep them in mind whilst generating your responses:
- You misunderstood/ignore the information provided at the beginning of the abstract.
- The edits you have made are not what we are aiming for, you tweaked a portion of the studies with non-significant findings, so there's no significant alternation of results occurring. Make sure your edit changes the main results of the studies, not trivial changes.
- Lack of inter-sentence consistency in the prompt
- You made edits as early as the first sentence. THe first few sentence are general knowledge and are not result of the study. So you shouldn't make any change in the beginning.
- Most of your edits contradict the conclusion. Make sure your changes do not contradit the conclusions or any part of the abstract.
- You only modified verbs the understanding of which does not require understanding of scientific concepts \& names of compounds, which makes the edits less likely to do wrong as long as reasons logically
- One of your edits contradicts all other edits.
- Your edit is inconsistent with the beginning of the sentence
- You failed to change the first part of the conclusion for consistency
- You missed out on one change.
- You misunderstood the purpose of the study. Although in the abstract it explicitly states the purpose of the study.

Below, you are given an abstract with its topic. Follow the instructions given to you and return the modified abstract. Remember to use double brackets to show the changes ([[original, modified ]] and keep the rest of the abstract unchanged. Also, pay attention to all the information you were given above as well as the common mistakes you have made before.

Abstract to edit:
Topic: \verb|<|abstract\_topic\verb|>|

Abstract: \verb|<|abstract\_to\_edit\verb|>|
"

\subsection*{Evaluations}
We tested human participants and LLMs on the BrainBench dataset. Both human experts and models were presented with two versions of the abstract, one with the actual results and one that was altered. The task was to determine which is which. Below, we detail how LLMs and human participants were tested.

\subsubsection*{Model Evaluation}
We tested LLMs by adapting the Eleuther AI Language Model Evaluation Harness framework \cite{gao_framework_2023}, which evaluates LLMs using a multiple choice setting. We presented LLMs with two versions of the abstracts from each test case separately. We prefixed each abstract with the prompt ``\textit{You are a neuroscientist with deep knowledge in neuroscience. Here is an abstract from a neuroscience publication:}'' and applied model-specific instruction templates where appropriate. We then measured the perplexity of both passages and used perplexity as the indicator of whether LLMs favor one abstract or the other.

Perplexity (PPL) is one of the most common metrics for evaluating large language models. Perplexity measures the degree of uncertainty of a model when generating a particular sequence of text. Formally, perplexity is defined as the exponentiated average negative log-likelihood of a tokenized sequence. If we have a tokenized abstract \( X = (x_0, x_1, \ldots, x_t) \), then the perplexity of \( X \), given a LLM parameterized by $\theta$ is,

\begin{equation}
    PPL(X) = \exp \left\{ -\frac{1}{t} \sum_{i}^{t} \log p_\theta (x_i | x_{<i}) \right\}
\end{equation}
where \( \log p_\theta (x_i | x_{<i}) \) is the log-likelihood of the \( i \)th token conditioned on the preceding tokens \( x_{<i} \) according to the LLM. Given both the original abstract \( X_{orig} \) and the altered abstract \( X_{alt} \), we followed the decision rule, 

\begin{equation}
    X_{chosen} = \begin{cases}
    X_{orig}, & \text{if } PPL(X_{orig}) < PPL(X_{alt}) \\
    X_{alt}, & \text{otherwise}
\end{cases}
\end{equation}
and evaluated the overall accuracy over the entire BrainBench accordingly.

\paragraph{Accuracy}
Accuracy is the primary metric for reporting LLM performance on BrainBench. A correct response was when the model produces a lower perplexity for the original abstract than the altered abstract.

\paragraph{Confidence calibration}
We used the absolute difference of perplexities of two versions of the abstract as a measure of model confidence. To assess the calibration of LLMs, we compared their accuracies with their confidence levels. First, we ranked and sorted model confidence across all test cases. Subsequently, we created 20 bins based on this sort. Within each bin, we calculated the mean accuracy. A well-calibrated model will exhibit a higher accuracy in bins associated with higher confidence rankings. We fit a linear regression model using the bin number as the independent variable and the mean accuracy of each bin as the dependent variable to evaluate calibration.

\paragraph{Performance correlation across LLMs}
We assessed the correlation in performance among different LLMs by examining how they rank the relative difficulty of test cases. To determine difficulty, we calculated the difference in perplexity between incorrect and correct abstracts for each test case. Intuitively, a large positive difference in the perplexity between incorrect and correct versions of an abstract should indicate that the test case is easy from the LLM's perspective. We calculated the Spearman correlation coefficient of these difficulty measures to assess the agreement between two LLMs. 

\paragraph{Integration analysis}
To investigate the extent to which LLMs can integrate broad context from abstracts, we conducted an experiment involving the removal of contextual information from BrainBench test cases. Following the same evaluation procedure as previously outlined for full abstract cases, we assessed the models using individual sentences extracted from abstracts containing at least one result alternation. In cases with multiple alternations, we computed the mean accuracy across these alternations as the final accuracy for the abstract. We then compared the level of performance degradation when LLMs were evaluated on full-length abstracts versus individual sentences where background and method information from the abstracts were removed.

In addition, we tested models using abstracts whose results (in terms of complete sentences) are randomly swapped from abstracts within the same neuroscience subfield. Importantly, in these ``swapped'' abstracts, the number of results remained consistent with the original. We applied the swapping to both original and altered abstracts and re-evaluated LLMs' performance.

\paragraph{LLM training data memorization analysis}
One concern regarding LLMs outperforming human experts on BrainBench is the possibility that LLMs were exposed to the original abstracts during their pre-training. If LLMs have simply memorized the training data, they would naturally assign lower perplexity scores to the correct abstracts.

To address this concern, we employed a common method from the literature to determine whether a given text is part of LLM's training data \cite{carlini_extracting_2021,nasr_scalable_2023}. This method involves calculating the zlib entropy and the perplexity ratio (Eq. \ref{eq:zlib_ppl_ratio}) of a text sequence to infer its membership status.

\begin{equation}
    ratio = \frac{ZLIB(X)}{PPL(X)}
\label{eq:zlib_ppl_ratio}
\end{equation}
Zlib entropy is computed using the zlib text compression algorithm \cite{gailly_zlib_2024}, which measures the level of uncertainty in a text when compressed. It's a data-agnostic way of evaluating text. On the other hand, LLM perplexity depends on the specific training data and thus is data-dependent. In general, if a piece of text surprises zlib but not LLM, it's likely part of the training data.

To conduct this test, we carefully chose data sources that are either known to be part of LLMs' pre-training or reasonably assumed to be excluded from it (refer to Table \ref{tab:models} and \ref{tab:data_sources}). We then applied zlib compression and LLM perplexity calculations to text samples from these selected sources.

In addition, we introduced the Gettysburg Address as a special anchor point to contrast with the zlib-perplexity ratio distribution across multiple data sources. This is because we expect the Gettysburg Address to exhibit a high zlib score due to its non-modern form of English, coupled with a low perplexity, given its likely frequent exposure during LLM pre-training.

Finally, we analyzed the Spearman correlation between the publication dates of the abstracts that make up BrainBench test cases against the test cases' difficulties to LLMs. This was to address the concern that early items are more likely to have a preprint or other precursor appear in the training set memorized by LLMs. If there was memorization, we would expect a negative correlation between publication date and difficulty. We determined difficulty by using the difference in perplexity between incorrect and correct abstracts for each test case.

\subsubsection*{Human Evaluation}
\paragraph{Participants}
We recruited 202 neuroscience experts via social media and an email newsletter. We excluded 31 participants for failing to answer both catch trials correctly, not providing confidence or expertise ratings during the entire experiment, and self-reported cheating. The remaining 171 participants consisted of 51 doctoral students, 43 faculty/academic staff, 43 postdoctoral researchers, 18 predoctoral students, 12 research scientists, and 4 classified as ``other". Participants' mean experience in neuroscience was 10.1 years. Participants identified as follows: 62.5\% male, 34.5\% female, and 0.6\% gender variant/non-conforming. The mean age was 35.2 years (\textit{SD = 9.4 years}).

\begin{figure}[H]
    \centering
    \includegraphics{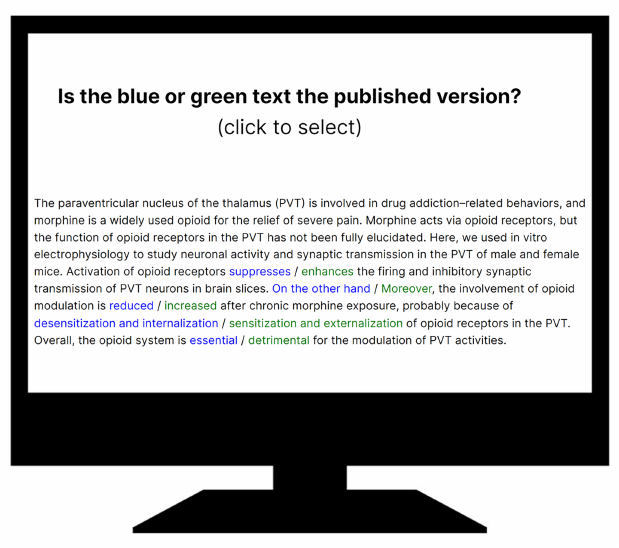}
    \caption{\textbf{Study test interface}: Participants were instructed to select which version of the abstract was the original by clicking on either blue or green text to select that set of options. Various test cases may have varying numbers of alternatives, but a single click will choose all options of the same color.}
    \label{fig:select_screen}
\end{figure}

\paragraph{Procedure}
First, participants were briefed on the experimental task and provided their informed consent to proceed to the experiment. Demographic information was then collected, including gender identity, age, country, current position, and years of experience in neuroscience research, broadly construed. Next, participants completed a practice trial using the same testing format as the actual test cases. This trial was used to familiarize participants with the format of the task with the screen proceeding only once participants had made the correct choice based on common sense. Following this, 9 test trials and 2 catch trials commenced, where participants selected one version of each trial abstract. Out of the 9 test trials, 6 were randomly sampled human-created test cases and 3 were randomly sampled from the pool of machine-created items. We ensured that each test case is sampled approximately an equal number of times across all participants. To achieve this, we maintained a global counter that keeps track of how frequently each test case has been used. As a result, the next participant's sample will always be drawn from those test cases that have been used less frequently. Notably, the number of alterations varies between test cases, but the design allowed a single click to automatically select between the two abstract options (see Fig. \ref{fig:select_screen}). Participants made one decision per test case, regardless of the number of alternations. 

Subsequently, participants were required to rate their confidence and expertise using slider bars. The confidence slider had a range from ``lower" on the left to ``higher" on the right, while the expertise slider spanned from ``not at all" on the left to ``very much so" on the right, both internally implementing a 1-100 scaling. Additionally, participants indicated whether they had encountered the study previously before proceeding to the next trial. Upon completing the 11 trials, participants were debriefed on which trials they got correct and were subsequently asked to indicate whether they engaged in any form of cheating during the study. We hosted the study entirely on the Gorilla platform \cite{anwyl-irvine_gorilla_2020}.

\paragraph{Exclusion criteria}
For participant selection and data analysis, we apply several exclusion criteria. First, individuals who failed to answer both catch trials correctly were not included in the data analyses. Second, participants who did not make adjustments to the sliders (i.e., expertise and confidence) during any of the trials were excluded. Additionally, trials where participants recognized the abstract content were omitted from the analysis. Furthermore, trials with reaction times less than 5 seconds were excluded. Lastly, participants who admitted to using external resources or engaging in cheating behaviors, as indicated by a checkbox in the debriefing form, were not considered in the final data analysis.

\paragraph{Performance correlation between humans and LLMs}
We assessed the agreement between humans and LLMs using a similar approach as we did when evaluating the correlation among LLMs. For LLMs, the procedure for determining item difficulty was identical to that described above.  For human experts, item difficulty was calculated as the mean accuracy for that item. Finally, the Spearman correlation of these difficulty measures was calculated to assess agreement.

\subsection*{Fine-tuning on neuroscience corpora}
The LLMs we considered had been pre-training on a diverse range of text corpora, including Internet sources, Wikipedia, books, code repositories, and arXiv papers. While these pre-trained models are designed to be versatile and capable of handling various tasks, our approach for creating BrainGPT involved enhancing base models with domain-specific expertise, specifically in neuroscience.

\begin{figure}[H]
    \centering
    \includegraphics[height=.6\textheight,width=.6\textwidth,keepaspectratio]{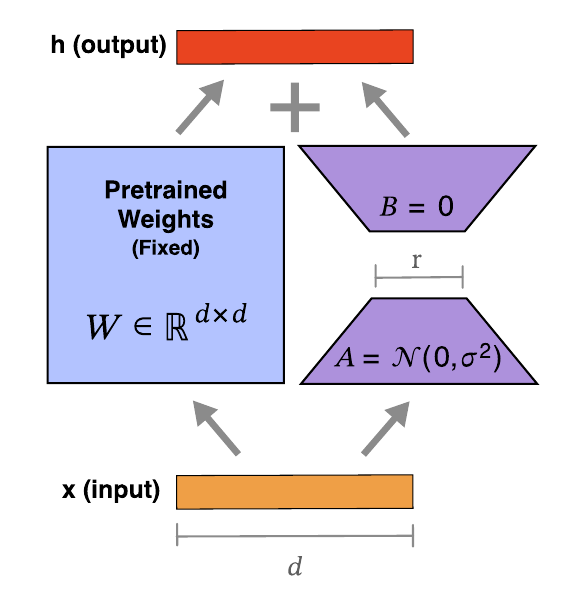}
    \caption{\textbf{Low-Rank Adaptation (LoRA) for parameter-efficient large language model (LLM) fine-tuning.} Image adapted from \cite{hu_lora_2021}. Training LLMs from scratch can be computationally expensive, especially when the model has many billions of parameters and the training data are massive. Ideally, one would be able to take advantage of a previously trained LLM (i.e., a base model) and build from it. Toward this end, various Parameter Efficient Fine-Tuning (PEFT) techniques have been proposed. Rather than retrain the LLM, these techniques preserve the LLM's original pre-trained weights and train a small subset of additional parameters that either enhance an existing capability in the LLM or introduce a new one. In our case, we tune a base LLM to the neuroscience literature.
    This tuning strategy significantly alleviates the computational burden while achieving comparable performance to training a new LLM from scratch. Among many PEFT approaches, we adopted LoRA in our study to fine-tune the 7-Billion parameter version of the LLaMa2 model on neuroscience abstracts. LoRA fixes the pretrained model weights and injects trainable rank decomposition matrices into each layer of the transformer. During training, the same input sequence (or intermediate layer output) $x$ is processed separately by the pretrained weights and the low-rank adaptation matrices ($A$ and $B$). The low-rank matrices possess the only trainable parameters in the model. The final output $h$ is computed as a coordinate-wise addition between the product of the pretrained weights and the adaptation matrices, which are further processed by subsequent layers.}
    \label{fig:lora}
\end{figure}

To accomplish this, we employed the Low-Rank Adaptation of Large Language Models (LoRA) technique (Fig. \ref{fig:lora}; \cite{hu_lora_2021}). LoRA efficiently extends the capabilities of general-purpose LLMs by introducing low-rank trainable parameters (referred to as ``adapters") into the existing model. This process effectively fine-tunes the model for downstream tasks without the need for prohibitively resource-intensive training of the entire model.

\paragraph{Training data}
We collected training data from PubMed for abstracts and PubMed Central Open Access Subset (PMC OAS) for full-text articles using the Entrez Programming Utilities (E-utilities) API and the pubget Python package, respectively. The data span publication dates from 2002 to 2022. For science general journals, we applied a keyword filter of ``Neuroscience" (see all sourced journals in Table \ref{table:journals}).

Our data extraction efforts yielded 332,807 abstracts and 123,085 full-text articles, totaling 1.3 billion tokens. We excluded figures and tables and randomly allocated 90\% of the data for training, reserving the remaining 10\% for validation.

\paragraph{Training details}
We fine-tuned Mistral-7B-v0.1 \mbox{\cite{jiang_mistral_2023}} using weights available on Huggingface (\href{https://huggingface.co/mistralai/Mistral-7B-v0.1{https://huggingface.co/mistralai/Mistral-7B-v0.1})}. We used a batch size of 1 and a chunk size of 2048. Training involved the use of the AdamW optimizer \cite{loshchilov_decoupled_2019} with a learning rate of 2e-5 and gradient accumulation steps set at 8. Two training epochs were performed, along with a warm-up step of 0.03 and a weight decay rate of 0.001. The learning rate was controlled using a cosine learning rate scheduler. LoRA adapters, characterized by a rank of 256, an alpha value of 512, and a dropout rate of 0.1, were applied after all self-attention blocks and fully-connected layers. This results in total 629,145,600 trainable parameters, roughly $8\%$ of the entire parameters of the base model. To optimize training performance, bf16 mixed precision training and data parallelism were employed. We used 4 Nvidia A100 (80GB) GPUs hosted on the Microsoft Azure platform. An epoch of training takes roughly 65 GPU hours.

\paragraph{Evaluation}
We tested the fine-tuned model on BrainBench using the same procedure as before. To verify the significance of performance improvement, we performed a paired t-test with respect to the perplexity of the correct options before and after fine-tuning.

\subsection*{Relative perplexities evaluated at token and sentence levels}
To gain further insights into how models solve BrainBench test cases, we broke down the change of relative perplexities over the sequence of tokens and sentences starting from the first alternation of results. Specifically, we looked at how often the model starts correctly at the first alternation and remains correct (Pos to Pos) or becomes incorrect (Pos to Neg) at the end. In addition, we categorized how often the model starts incorrectly but becomes correct (Neg to Pos) or remains incorrect (Neg to Neg).

\begin{figure}[H]
    \centering
    \includegraphics[scale=0.5]
    {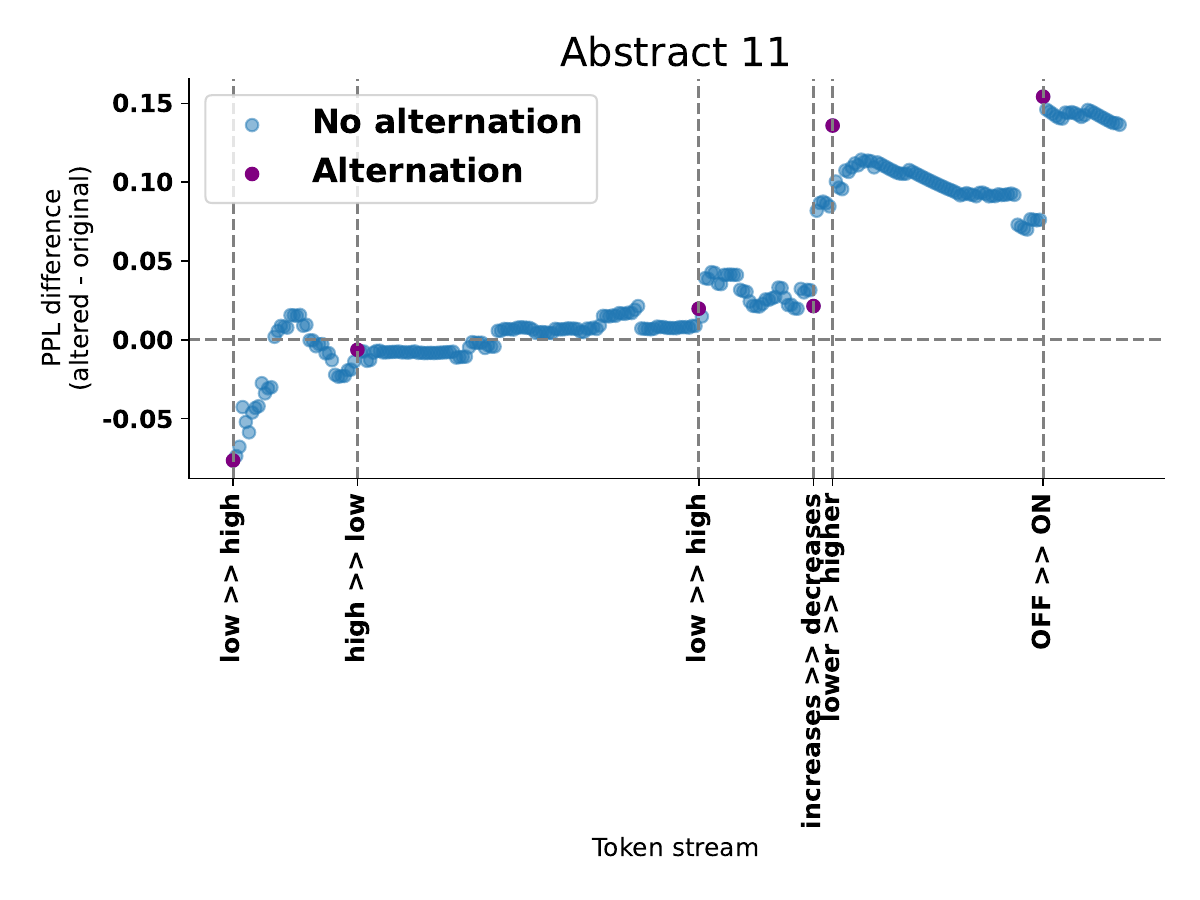}
    \caption{\textbf{Perplexity difference at token level.} Red dots indicate alternations. For this example, the model makes an incorrect response at the first alternation. As the model processes more tokens, the perplexity difference increases between the altered and original leading to a correct response at the end (Mistral-7B-v0.1).}
    \label{fig:token_11}
\end{figure}

\begin{figure}[H]
    \centering
    \includegraphics[scale=0.5]
    {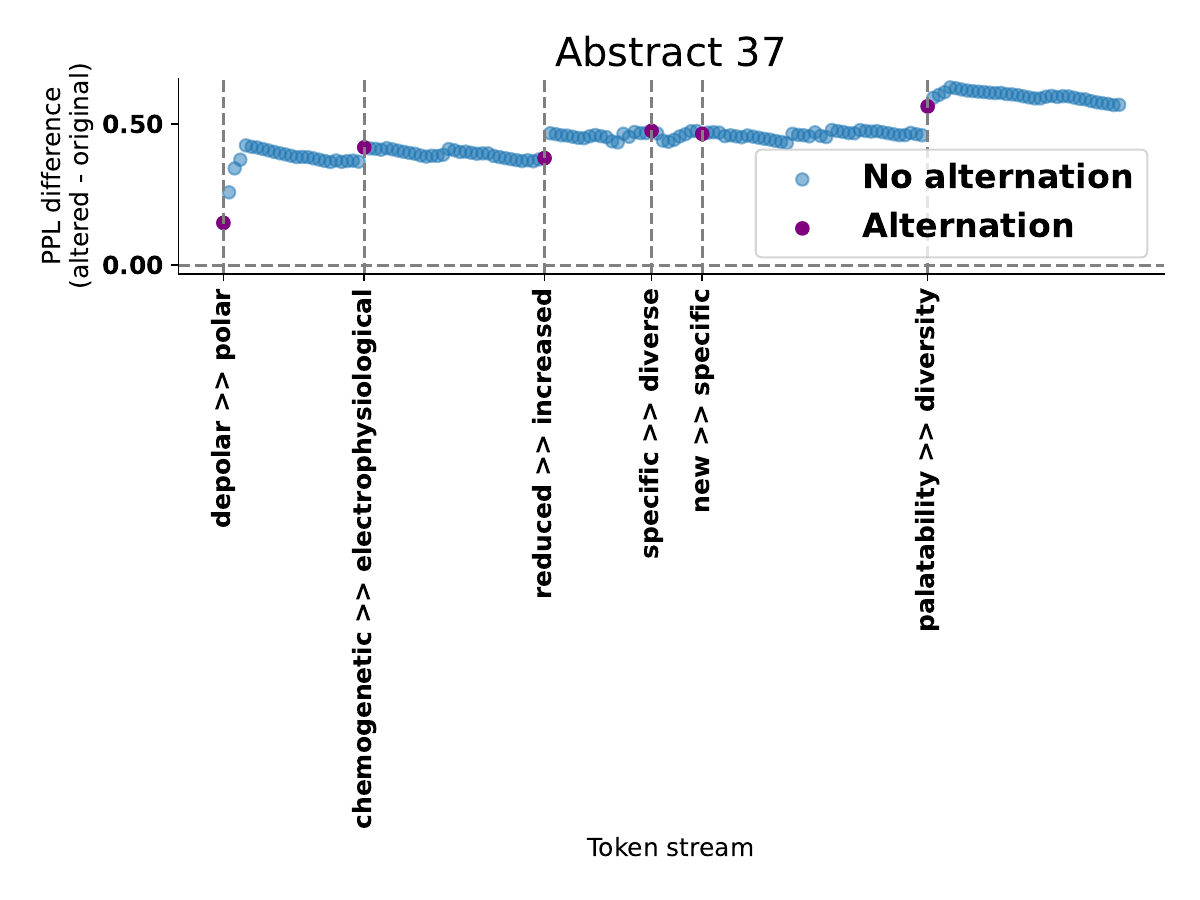}
    \caption{\textbf{Perplexity difference at token level.} Red dots indicate alternations. For this example, the model makes a correct response starting from the first alternation already and stays being correct throughout the rest of the tokens (Mistral-7B-v0.1).}
    \label{fig:token_37}
\end{figure}

\begin{figure}[H]
    \centering
    \includegraphics[scale=0.5]
    {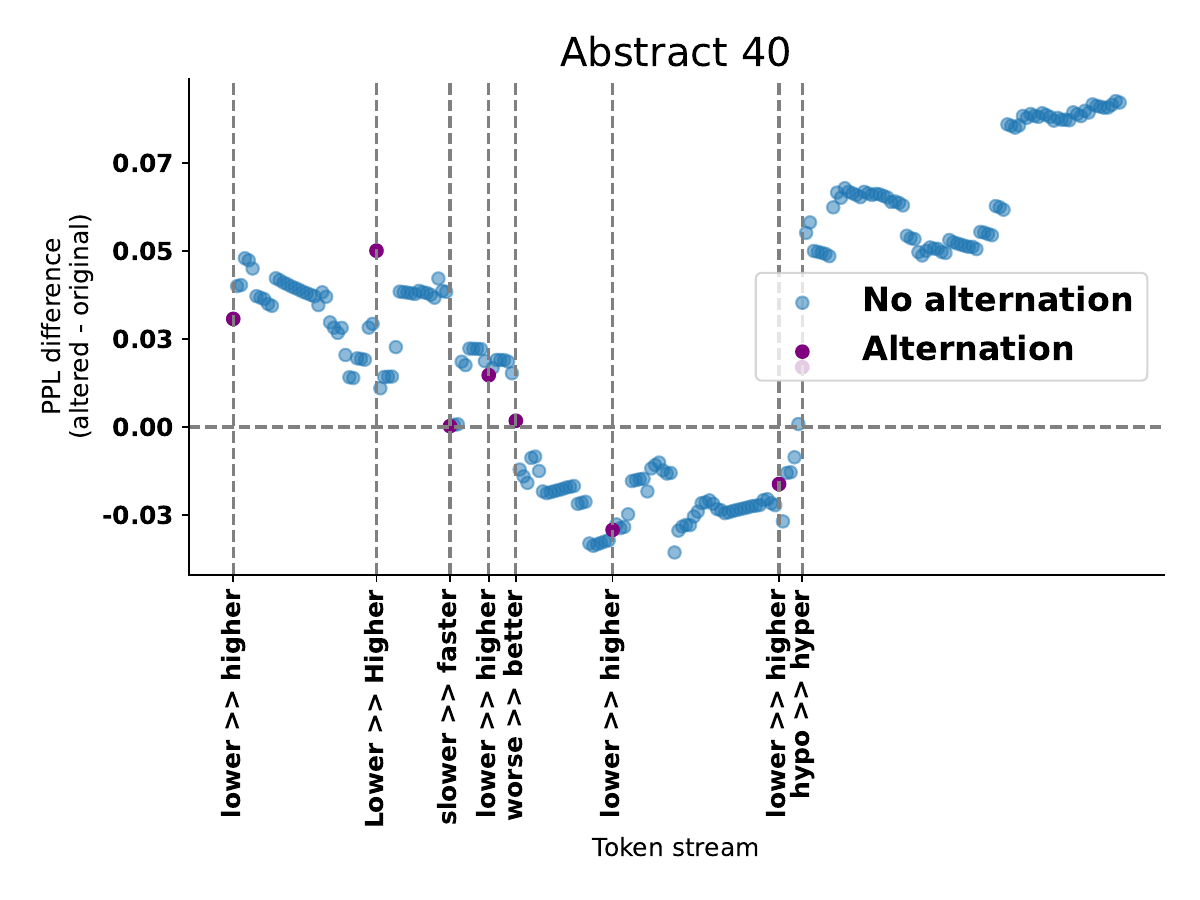}
    \caption{\textbf{Perplexity difference at token level.} Red dots indicate alternations. For this example, the model makes a correct response starting from the first alternation. The model switches to an incorrect response after having processed about half of the abstract. The model's response flips once again at the final alternation and makes a correct response at the end (Mistral-7B-v0.1).}
    \label{fig:token_40}
\end{figure}

Examples (Fig. \ref{fig:token_11}, \ref{fig:token_37}, \ref{fig:token_40}) show different patterns of relative perplexity changes between the altered and original abstracts over tokens. These examples demonstrate that the model does not always recognize the correct version at the first alternation, nor does it consistently remain in the correct (positive) or incorrect (negative) region across tokens.

\begin{figure}[H]
    \centering
    \includegraphics[scale=0.5]{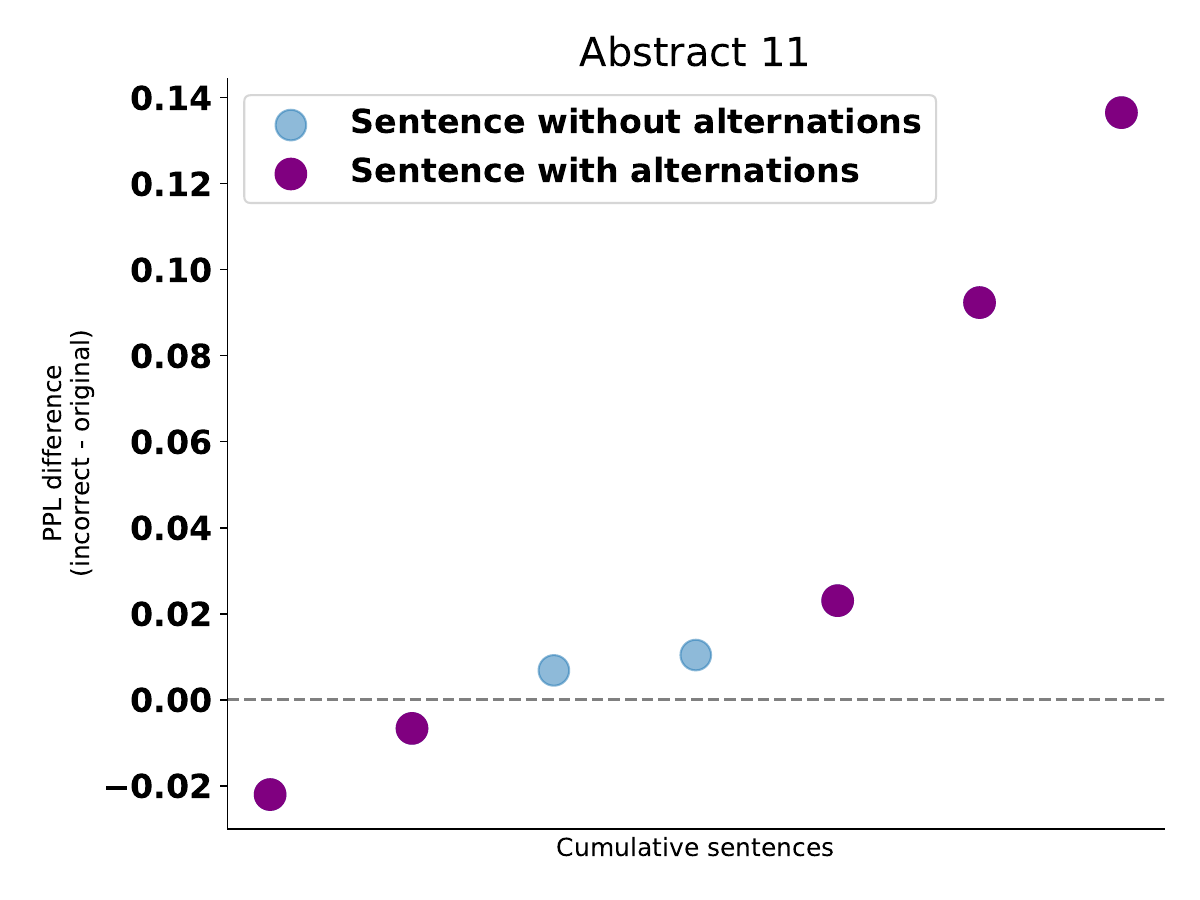}
    \caption{\textbf{Perplexity difference at sentence level.} Red dots indicate the current sentence contains at least one alternations.}
    \label{fig:sentence_11}
\end{figure}

\begin{figure}[H]
    \centering
    \includegraphics[scale=0.5]{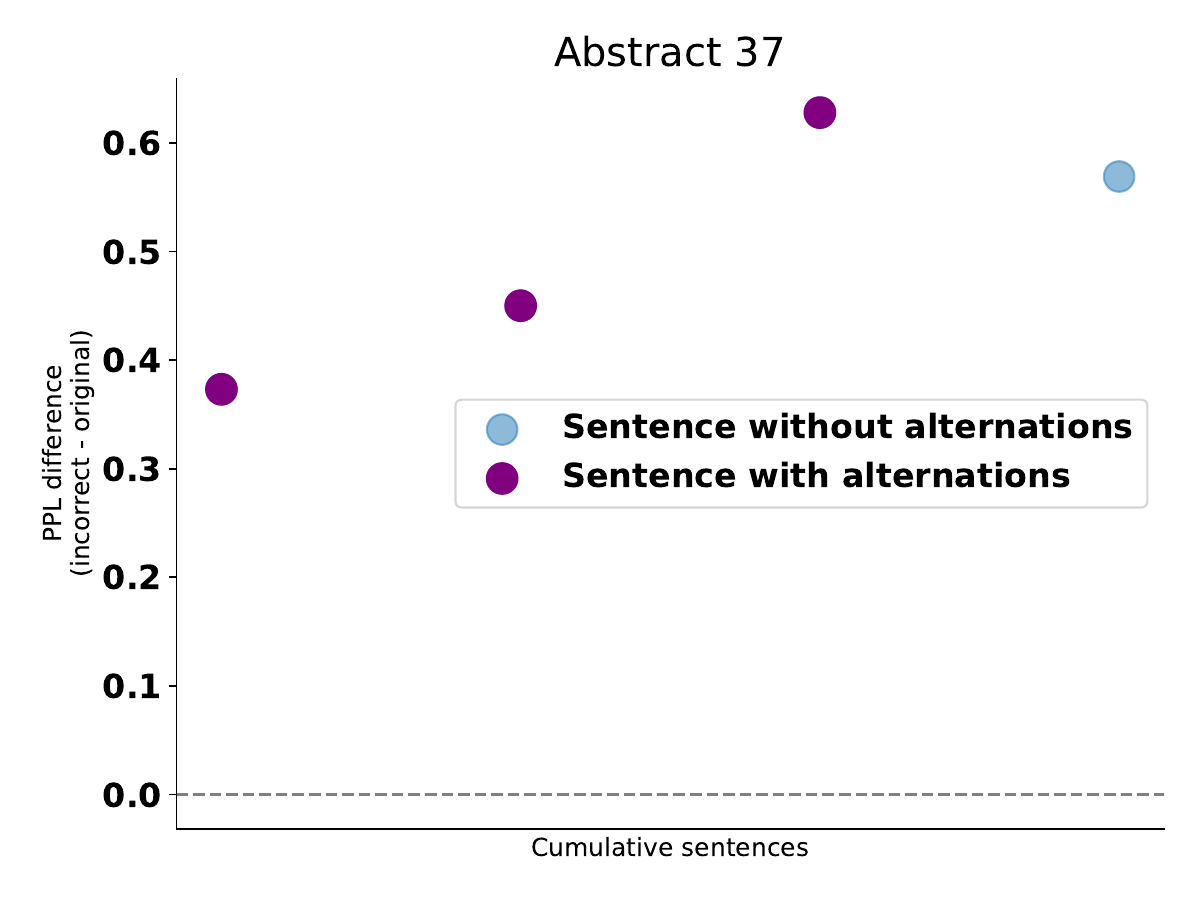}
    \caption{\textbf{Perplexity difference at sentence level.} Red dots indicate the current sentence contains at least one alternations.}
    \label{fig:sentence_37}
\end{figure}

\begin{figure}[H]
    \centering
    \includegraphics[scale=0.5]{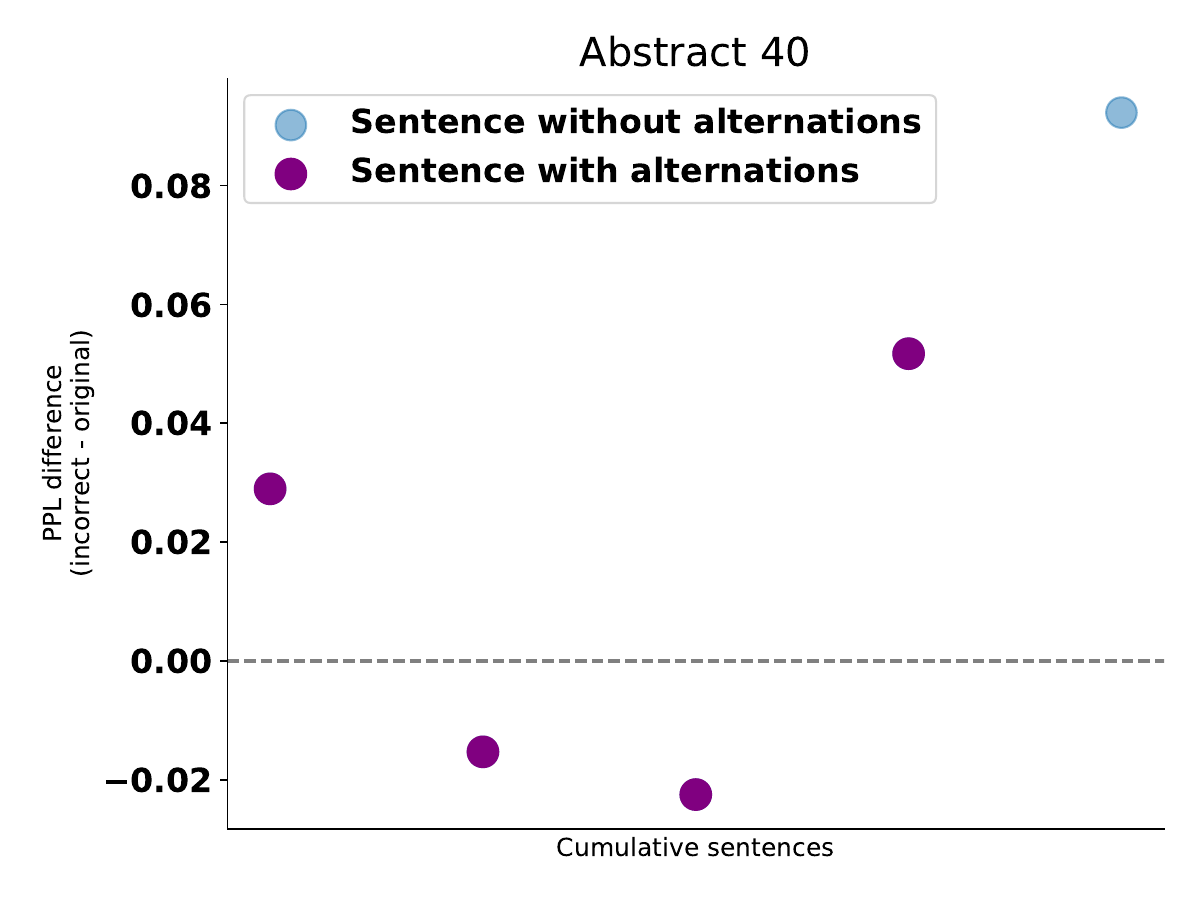}
    \caption{\textbf{Perplexity difference at sentence level.} Red dots indicate the current sentence contains at least one alternations.}
    \label{fig:sentence_40}
\end{figure}

\begin{figure}[H]
    \centering
    \includegraphics[width=\textwidth]{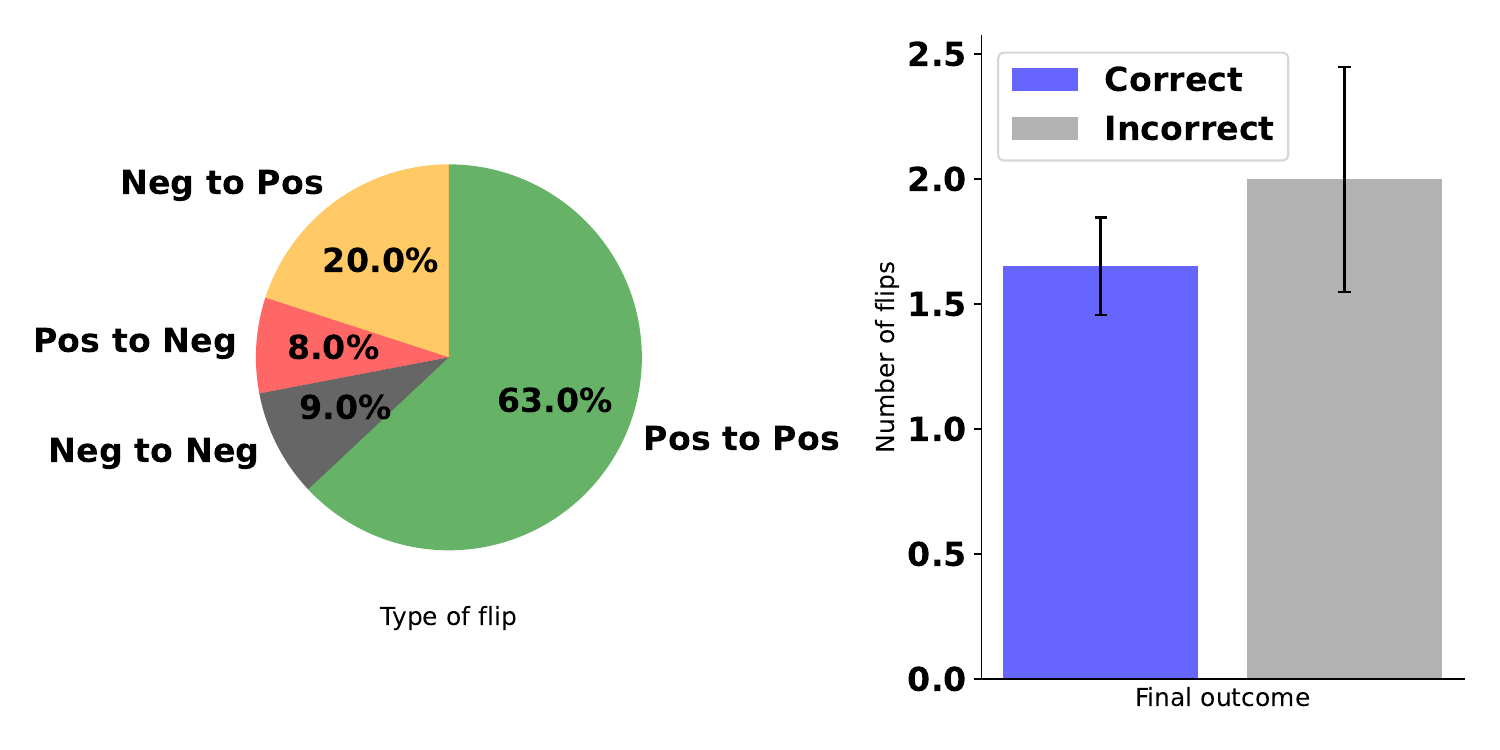}
    \caption{\textbf{Cumulative perplexity differences at token level}. Left panel shows proportions of different decision flips of a model. A flip is determined by comparing the cumulative perplexity difference between the incorrect and the original abstracts until the first diverging token and the entire abstract. For example, a ``Neg to Pos'' flip suggests the model has a negative perplexity difference at the first diverging token and has a positive perplexity difference at the final token. Right panel shows the average number of perplexity difference sign flips in between tokens among abstracts classified correctly and incorrectly by the model.}
    \label{fig:analysis_1_2_token}
\end{figure}

\begin{figure}[H]
    \centering
    \includegraphics[width=\textwidth]{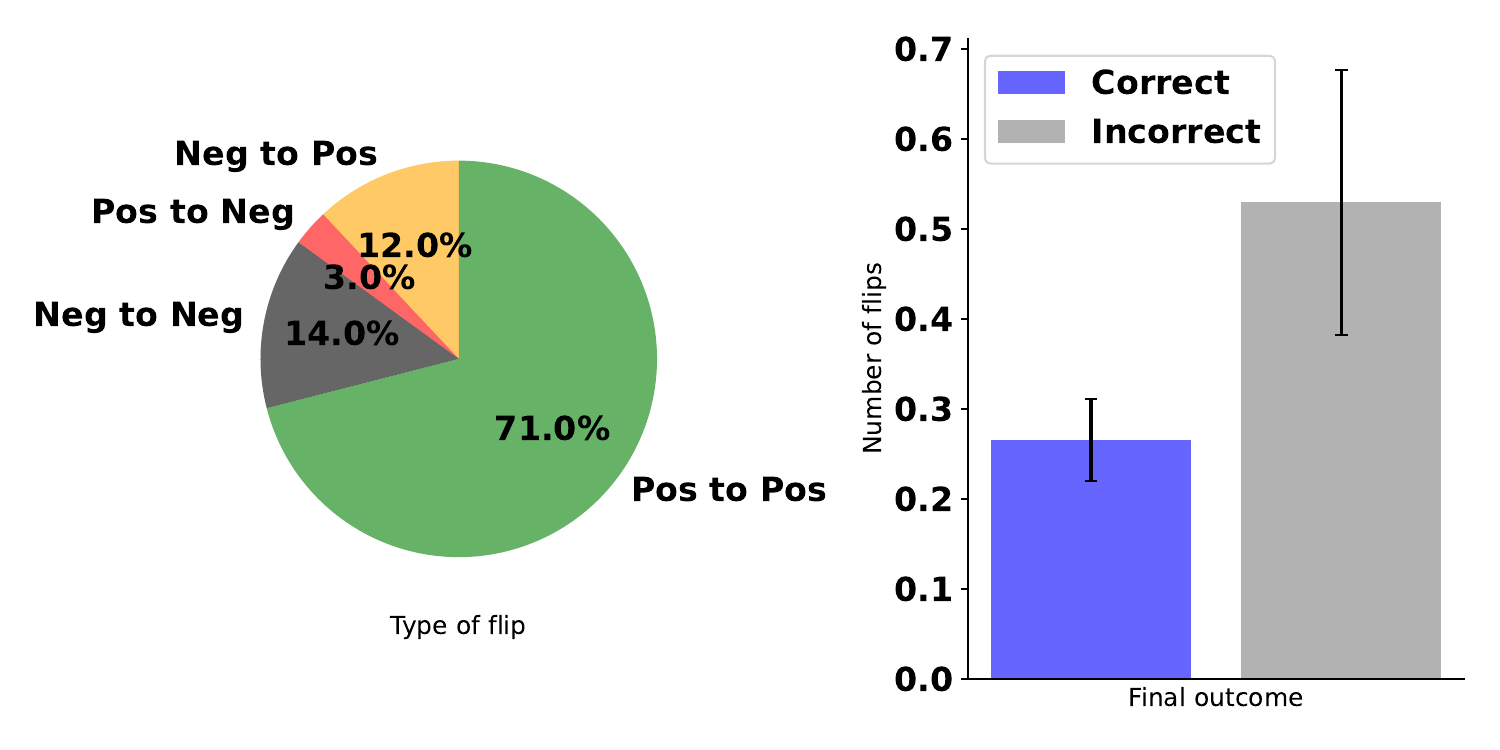}
    \caption{\textbf{Cumulative perplexity differences at sentence level}. Left panel shows proportions of different decision flips of a model. A flip is determined by comparing the cumulative perplexity difference between the incorrect and the original abstracts until the sentence that contains the first diverging token and the entire abstract. For example, a ``Neg to Pos'' flip suggests the model has a negative perplexity difference at the sentence that contains the first diverging token and has a positive perplexity difference at the final token. Right panel shows the average number of perplexity difference sign flips in between sentences among abstracts classified correctly and incorrectly by the model.}
    \label{fig:analysis_1_2_sentence}
\end{figure}

To gain a holistic view, we examined all test cases and categorized how the model solves a question. Specifically, we looked at how often the model starts correctly at the first alternation and remains correct (Pos to Pos) or becomes incorrect (Pos to Neg) at the end. We also categorized how often the model starts incorrectly but becomes correct (Neg to Pos) or remains incorrect (Neg to Neg). The corresponding results are shown in Fig. \ref{fig:analysis_1_2_token} (left panel). For more than half the abstracts, if the model is correct at the first alternation, it is likely to be correct at the end. Notably, $20\%$ of abstracts classified incorrectly at the first alternation are later classified correctly. We further evaluated beyond the first and final tokens by examining the number of sign flips of perplexity difference in between all tokens. As shown in Fig. \ref{fig:analysis_1_2_token}(right panel), the model exhibits fewer sign flips for abstracts eventually classified correctly than for those classified incorrectly. We performed the same analysis at the sentence level and observed similar patterns (Fig. \ref{fig:sentence_11}, \ref{fig:sentence_37}, \ref{fig:sentence_40}, \ref{fig:analysis_1_2_sentence}).

\subsection*{On the relationship between performance, model size and training data}
\label{SI:more}
\begin{figure}[H]
    \centering
    \includegraphics[width=.9\textwidth]{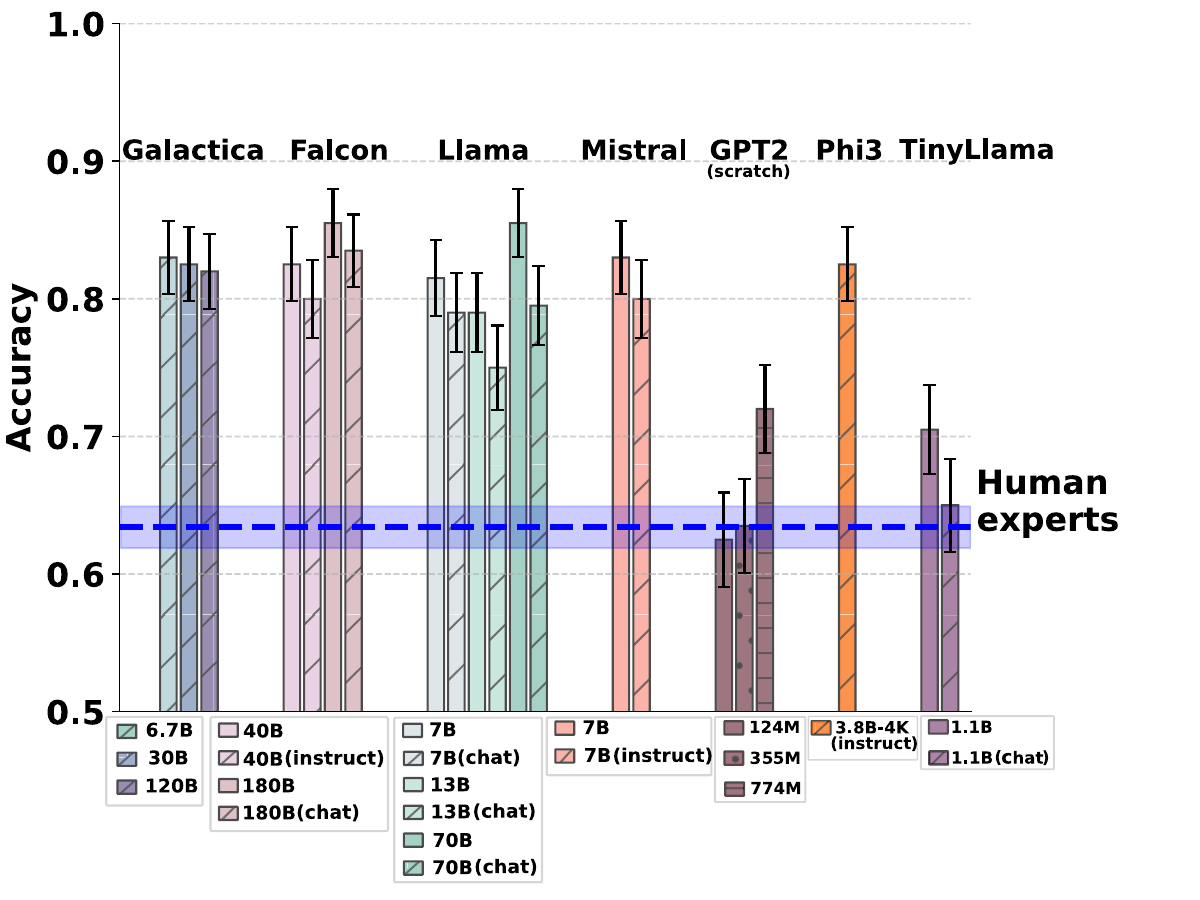}
    \caption{\textbf{BrainBench performance across models of varying sizes and training data.} Models with 7 billion parameters or more achieve similar results on BrainBench, while their instruct/chat fine-tuned variants perform systematically worse. The GPT-2 series (124M, 355M, 774M) trained entirely on the neuroscience literature from scratch shows progressively better results, matching or surpassing human performance and closing the gap to larger models. Phi3, with half the size of the 7B models, achieves competitive results likely due to its high-quality training data. In contrast, TinyLlama, with 1.1 billion parameters, lags behind on BrainBench. Overall, performance is influenced by both model size and the quality and relevance of the training data.}
    \label{fig:more}
\end{figure}

What factors might be driving LLMs' performance on BrainBench? As shown in Fig. \ref{fig:results_3_in_1}A, models of varying sizes starting from 7 billion parameters perform similarly, with instruct/chat fine-tuned models performing worse and larger models do not necessarily achieve higher accuracy. To gain better insights, we trained a much smaller model, GPT-2 (124 million parameters) from scratch, entirely on the same neuroscience data utilized in this study. We trained dedicated tokenizer on the same data. We trained GPT-2 for 5 epochs until convergence and tested it on BrainBench. The model achieved $63\%$ accuracy \cite{luo_matching_2024}, which is very close to human-level performance but falling behind on bigger models. To map out the picture more clearly, we further trained GPT-2 medium (355M) and GPT-2 large (774M) following the same training procedure. They both outperformed GPT-2 (124M) with GPT-2 large achieving a $72\%$ accuracy further reducing the gap to much larger models we tested (see full results in Fig. \ref{fig:more}). This pattern of result suggests that the size of the model likely plays a critical role in integrating deep knowledge to provide accurate responses.

In addition, we tested another small LLM, TinyLLama (1.1B; v1.0), as a comparison, and the model achieved $70.5\%$ (base) and $65.5\%$ (chat) on BrainBench; falling much behind the 7B models we tested in the original study. Further, we conducted testing on Phi-3 (3.8B), a very recent (April, 2024) model developed by Microsoft renowned for its competitive edge even against models more than twice its size. Phi-3 achieved an $82.5\%$ accuracy on BrainBench. We attribute this performance to the known high-quality training data curated by Microsoft.

In light of all these results, we posit that achieving superhuman performance on BrainBench likely stems from a synergy between model size and training set quality. The Phi-3 result raises intriguing questions about the existence of a scaling law in this context. It prompts exploration into whether there's a point of diminishing returns in model scaling and whether there exists a balance between model size, data quality and relevancy, and computational resources for optimal performance on BrainBench.

\newpage
\subsection*{Tables}
\begin{table}[H]
\centering
\begin{tabular}{|l|l|l|}
\hline
\textbf{Model Series} & \textbf{Pretraining Data Cutoff} & \textbf{Initial Release Date} \\ \hline
Llama2               & September 2022                   & July 2023                     \\ \hline
Galactica            & July 2022                        & November 2022                 \\ \hline
Falcon               & December 2022                    & May 2023                      \\ \hline
Mistral              & Unknown                          & September 2023                \\ \hline
\end{tabular}
\caption{Overview of LLMs Training Cutoff and Initial Release Dates.}
\label{tab:models}
\end{table}

\begin{table}[H]
\centering
\begin{tabular}{|l|l|}
\hline
\textbf{Data Source}        & \textbf{Model Training Set Inclusion} \\ \hline
RefinedWeb                  & Known to be in Falcon's training set \cite{almazrouei_falcon_2023}  \\ \hline
Arxiv (Jun-Dec 2021)                & Likely to be in Llama2's training set; known to be in Galactica's training set                                     \\ \hline
Arxiv (Jun-Dec 2023)                & Known to not be in Llama2 and Galactica's training set \cite{touvron_llama_2023,taylor_galactica_2022}                                     \\ \hline
Biorxiv (Jun-Dec 2021)              & Known to be in Galactica's training set \cite{taylor_galactica_2022}\\ \hline
Biorxiv (Jun-Dec 2023)              & Known to not be in Galactica's training set \\ \hline
Gettysburg Address          & Likely to be in all models' training set \\ \hline
\end{tabular}
\caption{Data Source Inclusion in AI Model Training Sets}
\label{tab:data_sources}
\end{table}

\begin{table}[H]
\centering
\begin{tabular}{@{}llll@{}}
\toprule
Model/Human                           & Coefficient           & Intercept             & Test Statistics        \\ \midrule
Galactica-6.7B                  & $1.69 \pm 0.04$       & $1.18 \pm 0.04$       & $t(4)=42.65$, $p<.001$ \\
Galactica-30B                   & $2.03 \pm 0.04$       & $1.11 \pm 0.02$       & $t(4)=54.55$, $p<.001$ \\
Galactica-120B                  & $2.32 \pm 0.12$       & $0.92 \pm 0.08$       & $t(4)=19.97$, $p<.001$ \\
Falcon-40B                      & $2.03 \pm 0.08$       & $1.09 \pm 0.06$       & $t(4)=24.64$, $p<.001$ \\
Falcon-40B (instruct)           & $2.50 \pm 0.09$       & $0.83 \pm 0.03$       & $t(4)=28.16$, $p<.001$ \\
Falcon-180B                     & $1.96 \pm 0.08$       & $1.36 \pm 0.06$       & $t(4)=25.90$, $p<.001$ \\
Falcon-180B (chat)              & $1.97 \pm 0.10$       & $1.02 \pm 0.07$       & $t(4)=18.99$, $p<.001$ \\
Llama-2-7B                      & $1.43 \pm 0.06$       & $1.32 \pm 0.06$       & $t(4)=24.53$, $p<.001$ \\
Llama-2-7B (chat)               & $1.39 \pm 0.09$       & $0.93 \pm 0.05$       & $t(4)=15.82$, $p<.001$ \\
Llama-2-13B                     & $1.92 \pm 0.07$       & $1.13 \pm 0.03$       & $t(4)=28.74$, $p<.001$ \\
Llama-2-13B (chat)              & $2.14 \pm 0.02$       & $0.59 \pm 0.03$       & $t(4)=122.38$, $p<.001$ \\
Llama-2-70B                     & $1.43 \pm 0.07$       & $1.63 \pm 0.09$       & $t(4)=21.27$, $p<.001$ \\
Llama-2-70B (chat)              & $2.30 \pm 0.14$       & $0.71 \pm 0.04$       & $t(4)=16.32$, $p<.001$ \\
Mistral-7B                      & $1.76 \pm 0.03$       & $1.28 \pm 0.03$       & $t(4)=55.71$, $p<.001$ \\
Mistral-7B (instruct)           & $1.90 \pm 0.08$       & $1.01 \pm 0.06$       & $t(4)=23.74$, $p<.001$ \\
Human experts                   & $0.01 \pm 0.00$       & $0.02 \pm 0.02$       & $t(4)=17.34$, $p<.001$ \\ \bottomrule
\end{tabular}
\caption{\textbf{Logistic regression fits.} For models, logistic regressions were fitted between perplexity differences of a test case and its correctness given a LLM. For human experts, logistic regressions were fitted between their confidence of a test case and its correctness. The significant positive correlations between model and human fits suggest that LLMs and humans are calibrated.}
\label{tab:lr_calibration}
\end{table}

\begin{table}[H]
\centering
\begin{tabular}{|p{\textwidth}|}
\hline
Nature, Cell, Cell Reports, eLife, Science Advances, Nature Communications, PNAS, The EMBO Journal,
Nature Neuroscience, Neuron, Brain, NeuroImage, Molecular Psychiatry, Journal of Neuroscience, Nature Reviews Neuroscience, Cerebral Cortex, Annals of Neurology, Human Brain Mapping, Epilepsia, Clinical Neurophysiology, Trends in Cognitive Sciences, Biological Psychiatry, Translational Psychiatry, Neuroscience and Biobehavioral Reviews, Neuropsychopharmacology, Alzheimer's and Dementia, NeuroImage: Clinical, Neurobiology of Aging, Trends in Neurosciences, Nature Reviews Neurology, Brain Stimulation, Frontiers in Neuroscience, Movement Disorders, Nature Human Behaviour, Frontiers in Neurology, Cortex, Journal of Alzheimer's Disease, Neurobiology of Disease, Biological Psychiatry: Cognitive Neuroscience and Neuroimaging, Brain Structure and Function, Pain, Frontiers in Human Neuroscience, eNeuro, Current Opinion in Neurobiology, European Journal of Neuroscience, Frontiers in Aging Neuroscience, Alzheimer's Research and Therapy, Journal of Neurology, Glia, Epilepsy and Behavior, Brain Imaging and Behavior, Journal of Neurophysiology, Sleep, Neuroscience, Neuropsychologia, Journal of Neural Engineering, Molecular Neurobiology, Frontiers in Cellular Neuroscience, Neuropharmacology, Alzheimer's and Dementia: Diagnosis, Assessment and Disease Monitoring, Journal of Neuroinflammation, Epilepsia Open, Acta Neuropathologica Communications, Frontiers in Neuroinformatics, Current Opinion in Behavioral Sciences, Developmental Cognitive Neuroscience, Frontiers in Molecular Neuroscience, Cerebellum, Journal of Cognitive Neuroscience, Network Neuroscience, Annual Review of Neuroscience, Progress in Neurobiology, Epilepsy Research, Molecular Autism, Journal of Comparative Neurology, Social Cognitive and Affective Neuroscience, Brain Topography, Hippocampus, Seizure: the journal of the British Epilepsy Association, Psychophysiology, Frontiers in Behavioral Neuroscience, Journal of Neurotrauma, Journal of Physiology, Frontiers in Neural Circuits, Neurobiology of Learning and Memory, Journal of Neural Transmission, Frontiers in Neuroanatomy, International Journal of Neuropsychopharmacology, Neuroscientist, Brain Sciences, Behavioural Brain Research, Experimental Neurology, Progress in Neuro-Psychopharmacology and Biological Psychiatry, Neurological Sciences, Neurotherapeutics, Neuroscience Letters, Current Opinion in Neurology, Journal of Neuroscience Methods, Journal of Neurochemistry, Neuromodulation, Molecular Neurodegeneration, Frontiers in Systems Neuroscience, Sleep Medicine Reviews, Brain and Behavior, Brain Research, Neurorehabilitation and Neural Repair, Autism Research.
\\ \hline
\end{tabular}
\caption{\textbf{Journals used for LoRA fine-tuning.} Abstracts and full articles published between 2002 and 2022 sourced from the above journals were used as the training set for LoRA fine-tuning.}
\label{table:journals}
\end{table}

\begin{table}
\centering
\begin{tabular}{|l|p{10cm}|}
\hline
Tasks & Authors \\\hline
Primary writing & \begin{tabular}[t]{@{}l@{}}Xiaoliang Luo, Bradley C. Love\end{tabular} \\
\hline
Test case creation & \begin{tabular}[t]{@{}l@{}}Bati Yilmaz, Isil Poyraz Bilgin, Anton Pashkov\\Tereza Okalova, Alexandra O. Cohen, Felipe Yáñez\\Elkhan Yusifov, N Apurva Ratan Murty, Kangjoo Lee\\Valentina Borghesani, Sepehr Razavi, Justin M. Ales\\Rui Mata, Michael Gaebler, Guiomar Niso\\Leyla Loued-Khenissi, Anna Behler, Kaustubh R. Patil \\Mikail Khona, Roberta Rocca, Kevin K. Nejad\\Alessandro Salatiello, Jonathan Nicholas\\ Daniele Marinazzo, Chloe M. Hall\\Pui-Shee Lee, Sebastian Musslick\\Nicholas E. Myers, Jennifer K Bizley, Sherry Dongqi Bao\\Nianlong Gu, Jessica Dafflon, Bradley C. Love\end{tabular} \\
\hline
Quality control & \begin{tabular}[t]{@{}l@{}}Bati Yilmaz, Kevin K. Nejad, Daniele Marinazzo\\Pui-Shee Lee, Nianlong Gu, Chloe M. Hall\\Kangjoo Lee, Sebastian Musslick, Akilles Rechardt\\N Apurva Ratan Murty, Ilia Sucholutsky, Guiomar Niso\\Isil Poyraz Bilgin, Rui Mata, Tereza Okalova\\Mikail Khona, Justin M. Ales, Michael Gaebler\\Elkhan Yusifov, Leyla Loued-Khenissi, Jessica Dafflon\\Jonathan Nicholas, Felipe Yáñez, Roberta Rocca\\Valentina Borghesani, Sherry Dongqi Bao\\Alexandra O. Cohen, Alessandro Salatiello\\Xiaoliang Luo, Bradley C. Love\end{tabular} \\
\hline
GPT-4 case creation & \begin{tabular}[t]{@{}l@{}}Kevin K. Nejad, Xiaoliang Luo, Bradley C. Love\end{tabular} \\
\hline
Human-machine teaming & \begin{tabular}[t]{@{}l@{}}Felipe Yáñez, Xiaoliang Luo, Bradley C. Love\end{tabular} \\
\hline
LoRA fine tuning & \begin{tabular}[t]{@{}l@{}}Guangzhi Sun, Xiaoliang Luo, Bradley C. Love\end{tabular} \\
\hline
Model evaluation & \begin{tabular}[t]{@{}l@{}}Xiaoliang Luo, Guangzhi Sun, Martin Ferianc\\Bradley C. Love\end{tabular} \\
\hline
Building the experiment & \begin{tabular}[t]{@{}l@{}}Akilles Rechardt, Xiaoliang Luo, Bradley C. Love\end{tabular} \\
\hline
Data analysis & \begin{tabular}[t]{@{}l@{}}Xiaoliang Luo, Akilles Rechardt, Bradley C. Love\end{tabular} \\
\hline
Conceptualization and strategy & \begin{tabular}[t]{@{}l@{}}Xiaoliang Luo, Bradley C. Love\end{tabular} \\
\hline
Figure creation & \begin{tabular}[t]{@{}l@{}}Xiaoliang Luo, Bati Yilmaz, Isil Bilgin\\Chloe M. Hall, Guiomar Niso, Bradley C. Love\end{tabular} \\
\hline
Useful input and suggestions on project & \begin{tabular}[t]{@{}l@{}}All authors\end{tabular} \\
\hline
Commenting and editing on the manuscript & \begin{tabular}[t]{@{}l@{}}All authors\end{tabular} \\
\hline
\end{tabular}
\caption{\textbf{Author contributions breakdown.} For each task, authors listed toward the beginning of the list contributed more.}
\label{table:contributions}
\end{table}

\newpage
\subsection*{Figures}

\begin{figure}[H]
    \centering
\includegraphics[height=.6\textheight,width=.6\textwidth,keepaspectratio]{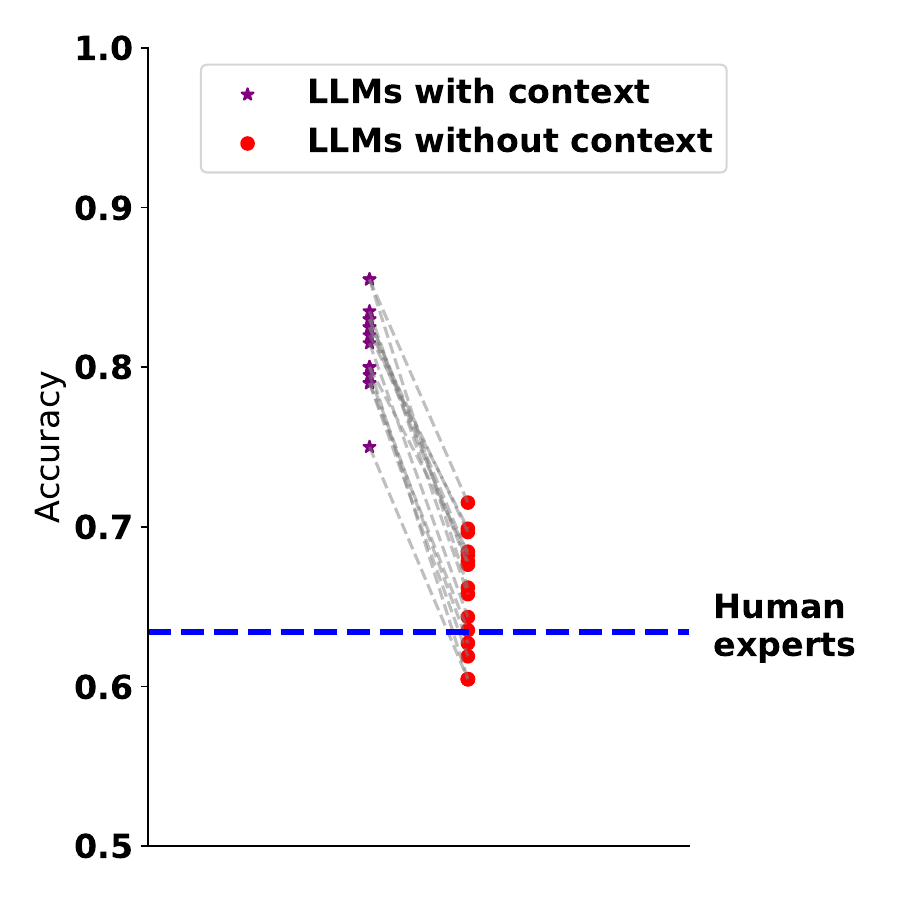}
    \caption{\textbf{LLM integrates contextual information to succeed on BrainBench}. The removal of background and method sections from abstracts, with an evaluation based solely on individual sentences and result alternations, significantly impairs the performance of LLMs on BrainBench. LLMs' superior performance appears to arise from integrating information across the abstract.
}
    \label{fig:iso}
\end{figure}

\begin{figure}[H]
    \centering
    \includegraphics[height=.6\textheight,width=.6\textwidth,keepaspectratio]{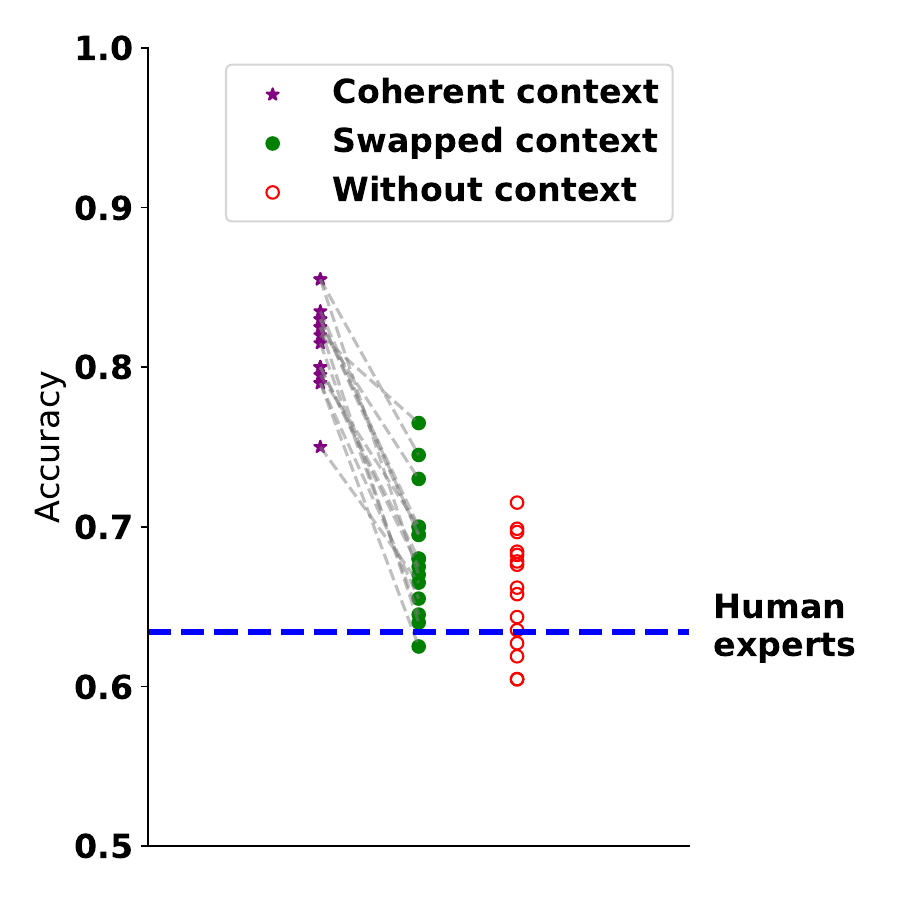}
    \caption{\textbf{LLM performance with coherent, swapped and no context.} There is a significant performance decline when evaluating with coherent versus swapped contexts, with further degradation when context is absent.}
    \label{fig:swap}
\end{figure}

\begin{figure}[H]
    \centering
\includegraphics[height=\textheight,width=\textwidth,keepaspectratio]{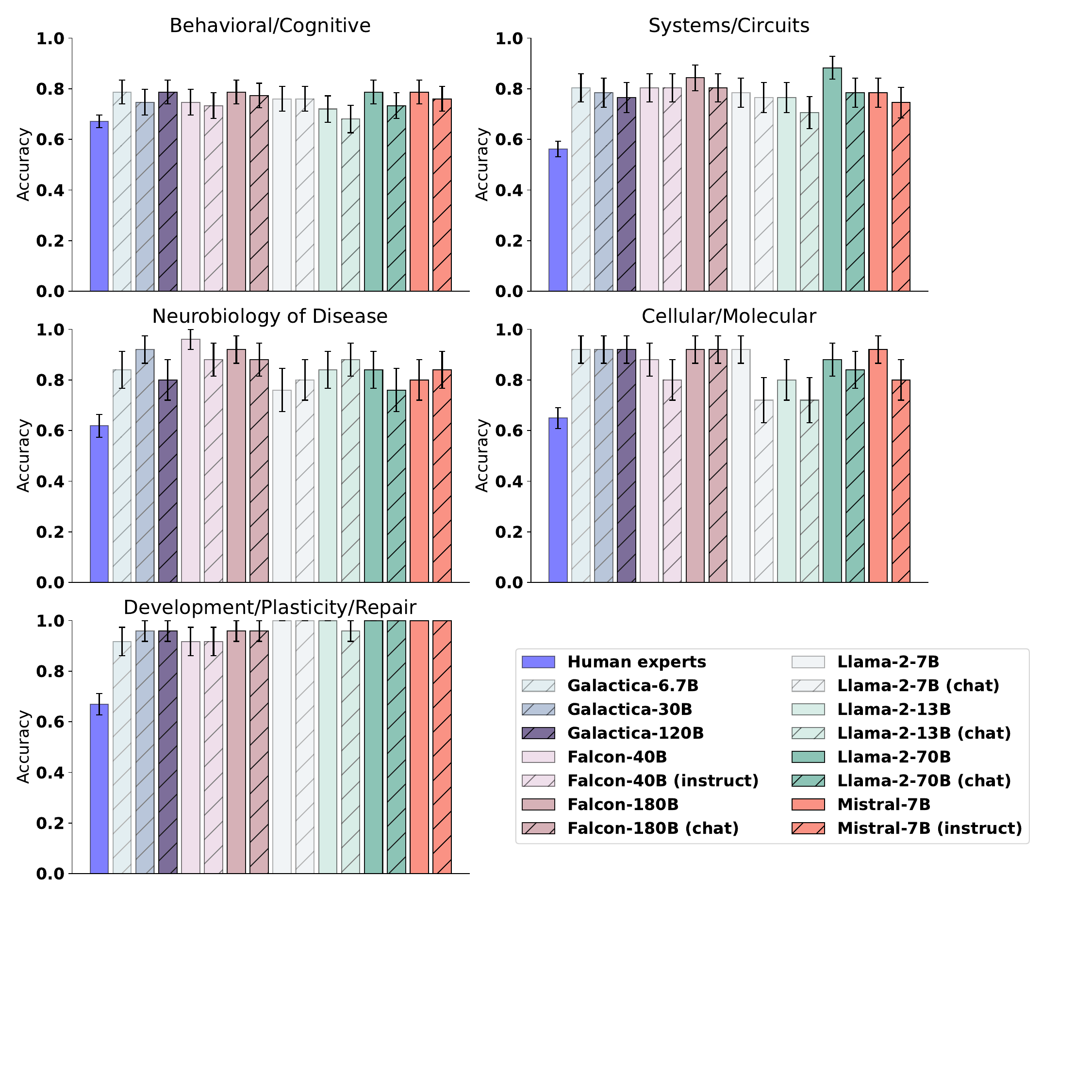}
    \caption{\textbf{BrainBench performance by subfield.} Accuracy on BrainBench by subfields of neuroscience for human experts and LLMs.}
    \label{fig:breakdown_subfield}
\end{figure}

\begin{figure}[H]
    \centering
    \includegraphics[height=.7\textheight,width=.7\textwidth,keepaspectratio]{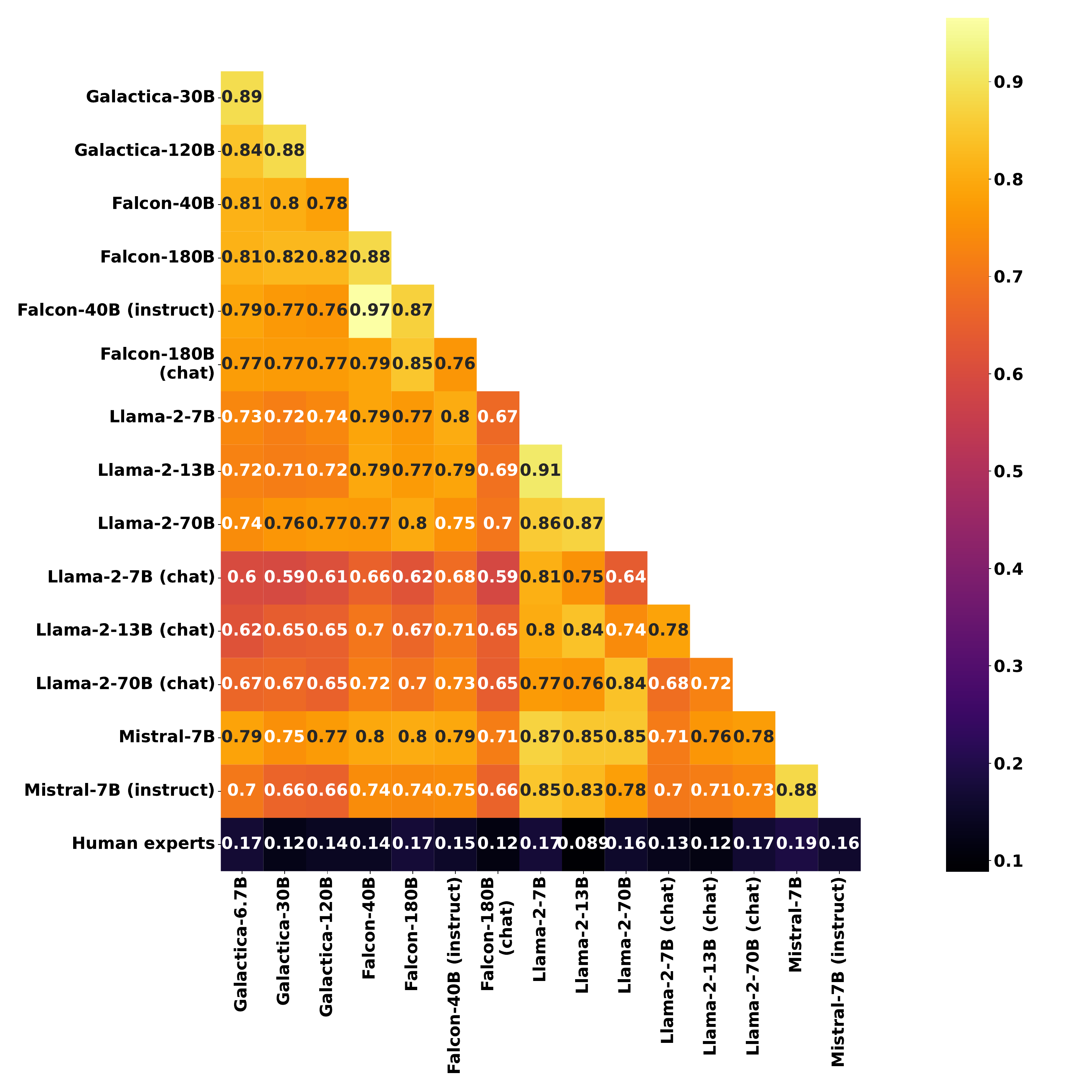}
    \caption{\textbf{Item difficulty correlation among LLMs and human experts.} For LLMs, difference in perplexity between incorrect and correct abstracts is used to determine the relative difficulty of test cases. Mean accuracy is used for human experts. Spearman correlation is calculated for these difficulty measures. LLMs have an average correlation of $0.75$ ($\pm0.08$) whereas human experts have an average of $0.15$ ($\pm0.03$) with LLMs.}
    \label{fig:error_correlation_heatmap}
\end{figure}

\begin{figure}[H]
    \centering    \includegraphics[height=\textheight,width=\textwidth,keepaspectratio]{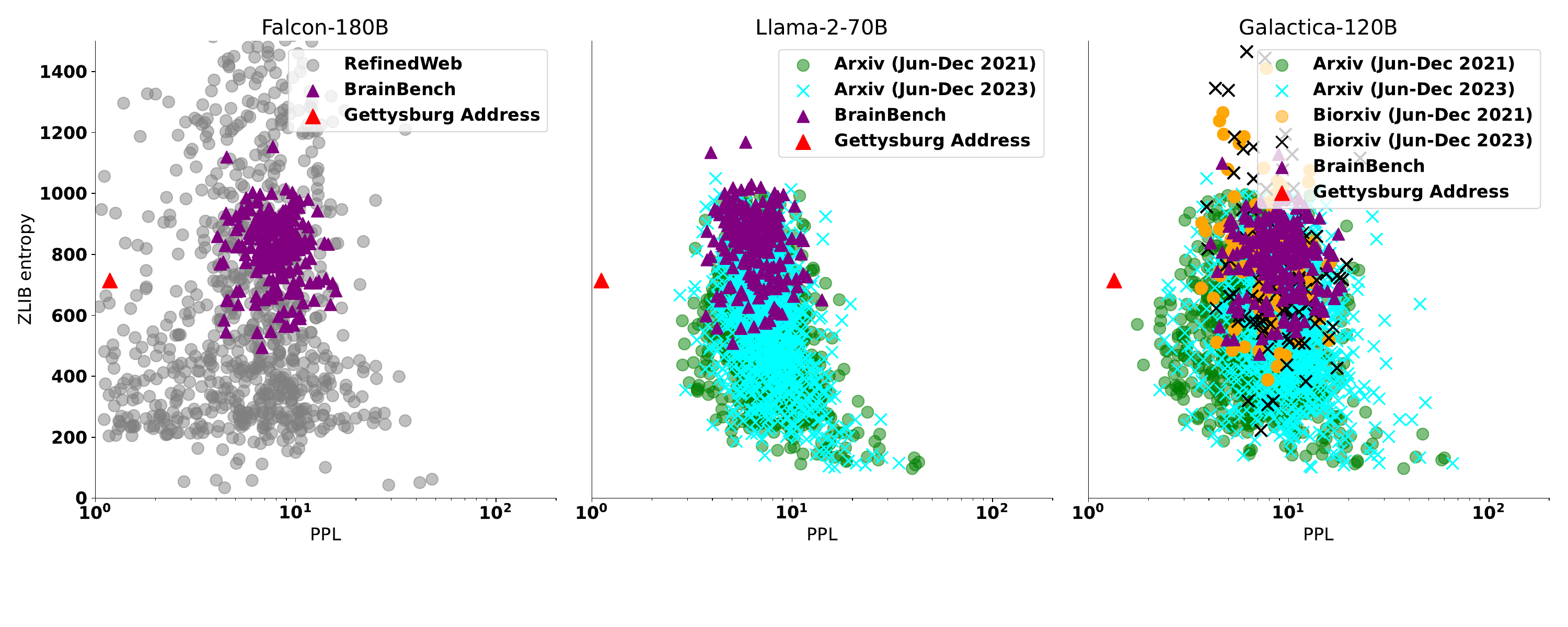}
    \caption{\textbf{Memorization analysis.} Analyzing zlib entropy and perplexity ratio in text samples from known data sources, both included and excluded from LLMs training data, reveals no distinct signature for pre-trained data sources compared to those outside the training set. Although certain training data that occurs multiple times in the training set, like the Gettysburg Address, show signs of memorization, low-frequency training data show no sign of memorization. We analyzed the three largest LLMs we evaluated as they are most capable of memorizing data.}
    \label{fig:memorisation}
\end{figure}

\begin{figure}[H]
    \centering
    \includegraphics[width=\textwidth]{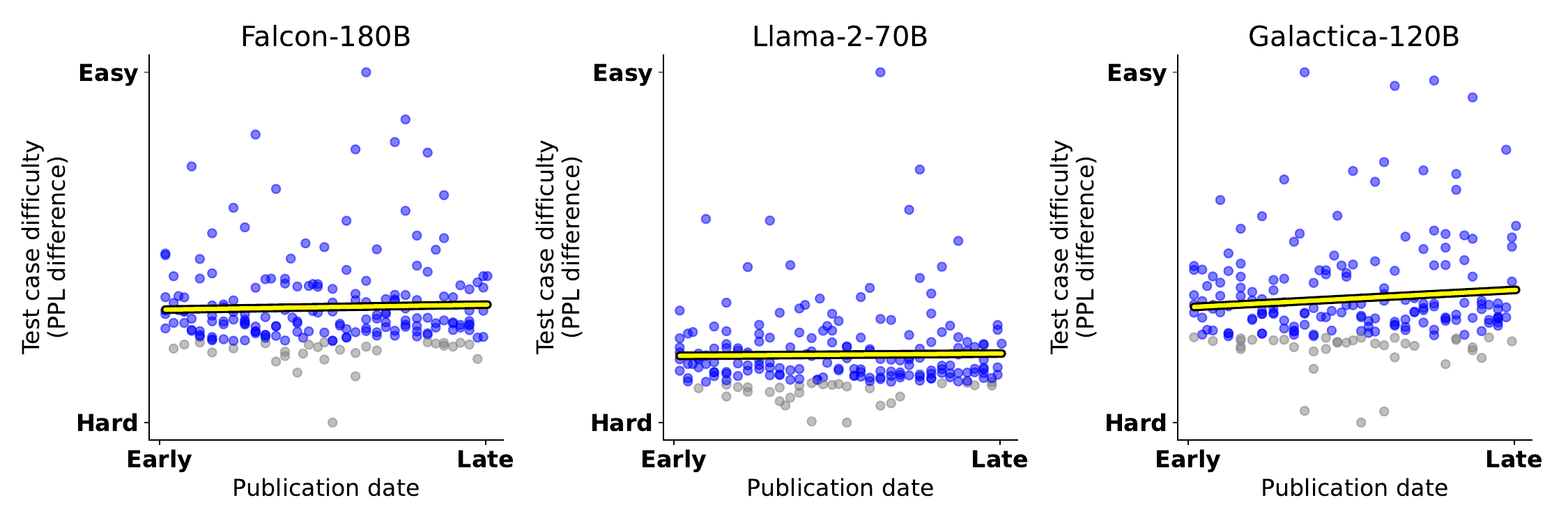}
    \caption{\textbf{Spearman correlation between test cases' publication date and difficulty.} No correlation was found between the publication dates of BrainBench test cases and their difficulties in terms of perplexity differences to LLMs (Falcon-180B, $r(198)=.00$, $p=.97$; Llama-2-70B, $r(198)=-.03$, $p=.65$; Galactica-120B, $r(198)=.03$, $p=.65$). Blue dots are test cases LLMs correctly classified; gray dots are test cases misclassified by LLMs.}
    \label{fig:pubdates_difficulty_correlation}
\end{figure}

\begin{figure}[H]
    \centering    \includegraphics[width=.5\textwidth]{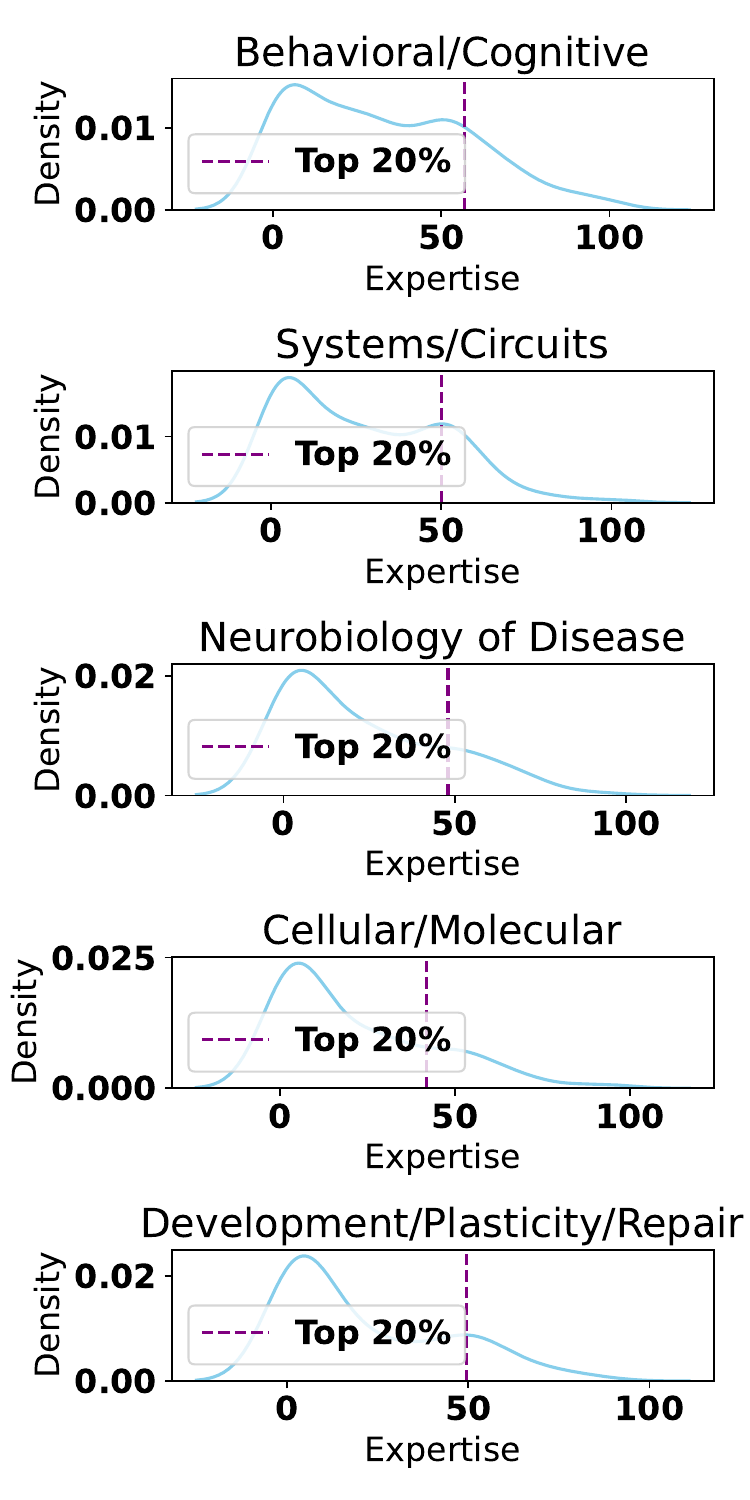}
    \caption{\textbf{Expertise distribution across subfields.} The distribution of expertise are generally similar across subfields. The top 20\% experts report expertise around or above 50/100.}
    \label{fig:expertise_distr_per_subfield}
\end{figure}

\begin{figure}[H]
    \centering
    \includegraphics[width=\textwidth]{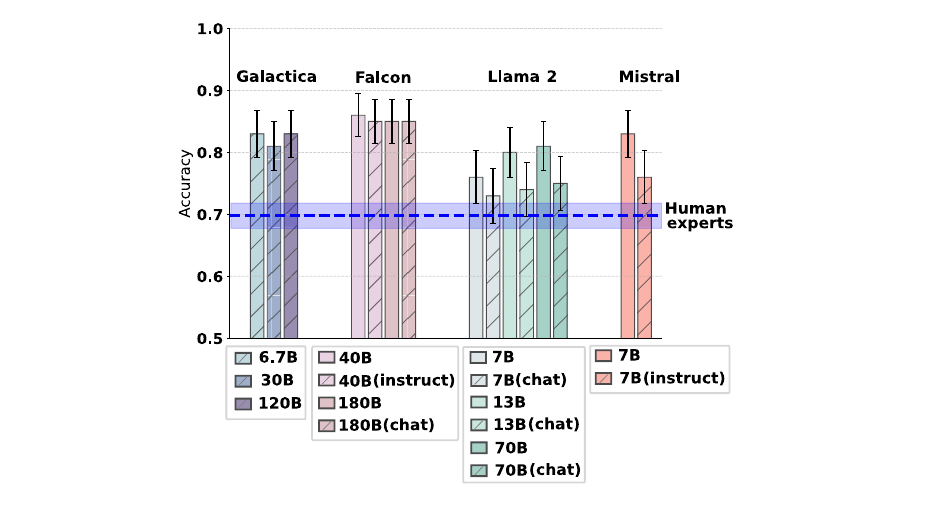}
    \caption{\textbf{LLMs outperformed human experts on BrainBench (using GPT-4 created test cases).}
    Base versions of models outperformed chat and instruct versions, which were tuned to be conversational with humans.}
    \label{fig:overall_accuracy_model_vs_human_machineCreated}
\end{figure}

\begin{figure}[H]
    \centering
    \includegraphics[height=.8\textheight,width=.8\textwidth,keepaspectratio]{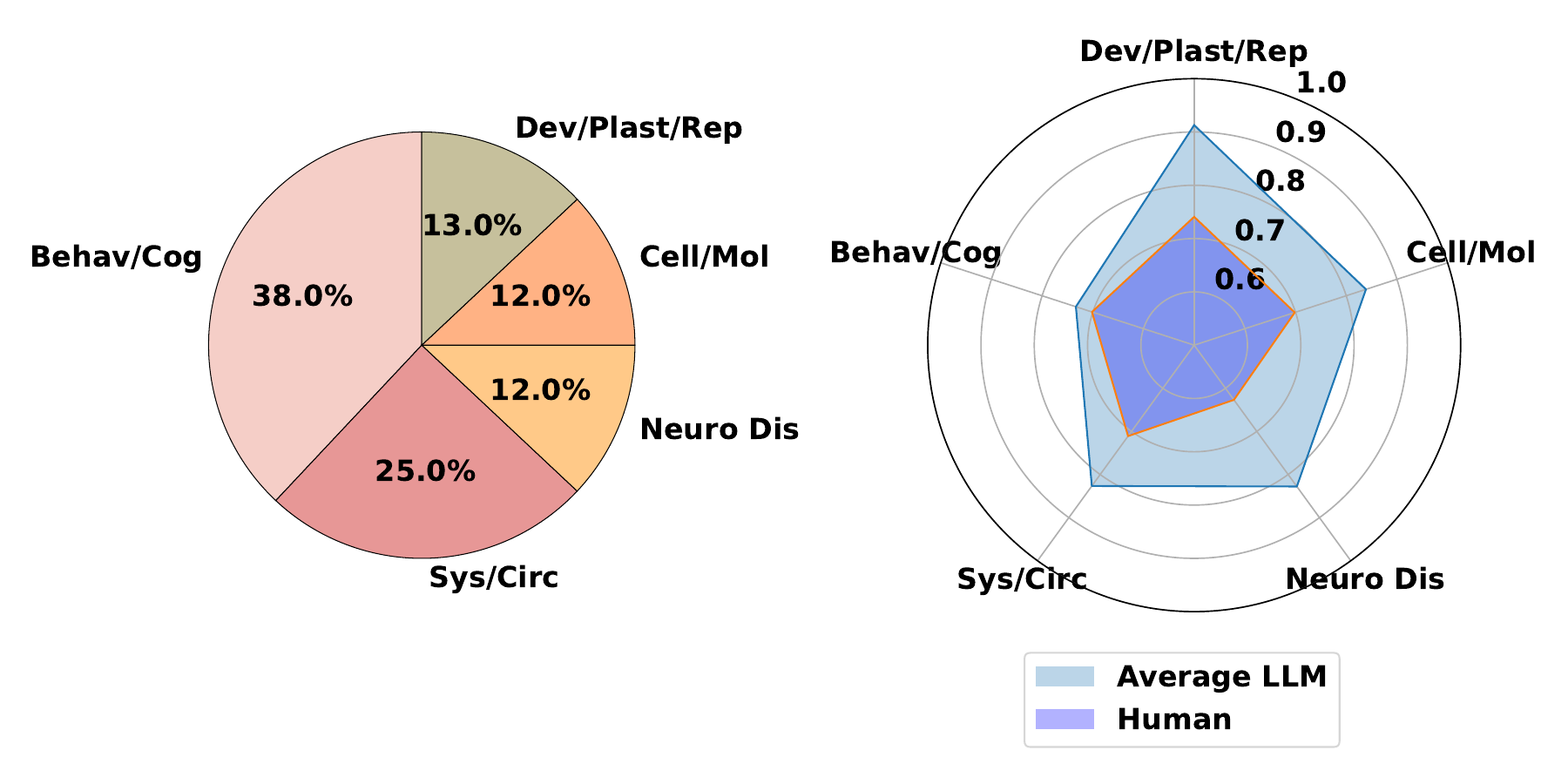}
    \caption{\textbf{BrainBench performance breakdown by subfields of neuroscience (using GPT-4 created test cases).} The distribution of test cases across neuroscience subfields roughly mirrors the distribution of articles in the Journal of Neuroscience with Behavior/Cognitive overrepresented. The mean performance of 15 LLMs and human experts is shown. LLMs outperformed human experts in all subfields.}
    \label{fig:subfields_machineCreated}
\end{figure}

\begin{figure}[H]
    \centering
    \includegraphics[width=.7\textwidth]{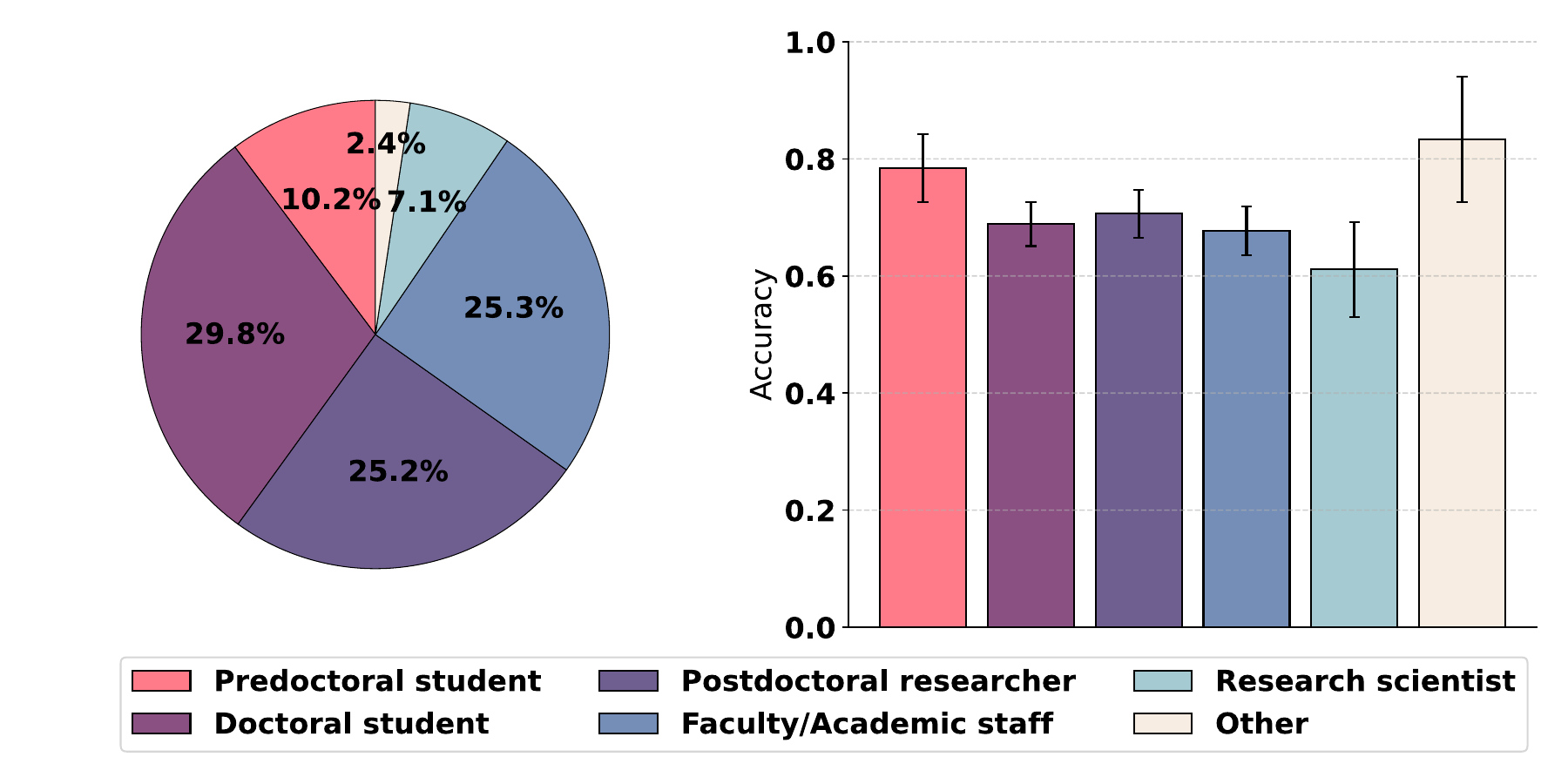}
    \caption{\textbf{BrainBench performance of human experts belong to different self-reported career categories (using GPT-4 created test cases).} Majority of the participants were doctoral students,
    postdoctoral researchers and faculty/academic staff.}
    \label{fig:acc_researcher_type_machineCreated}
\end{figure}

\begin{figure}[H]
    \centering
    \includegraphics[height=.7\textheight,width=.7\textwidth,keepaspectratio]{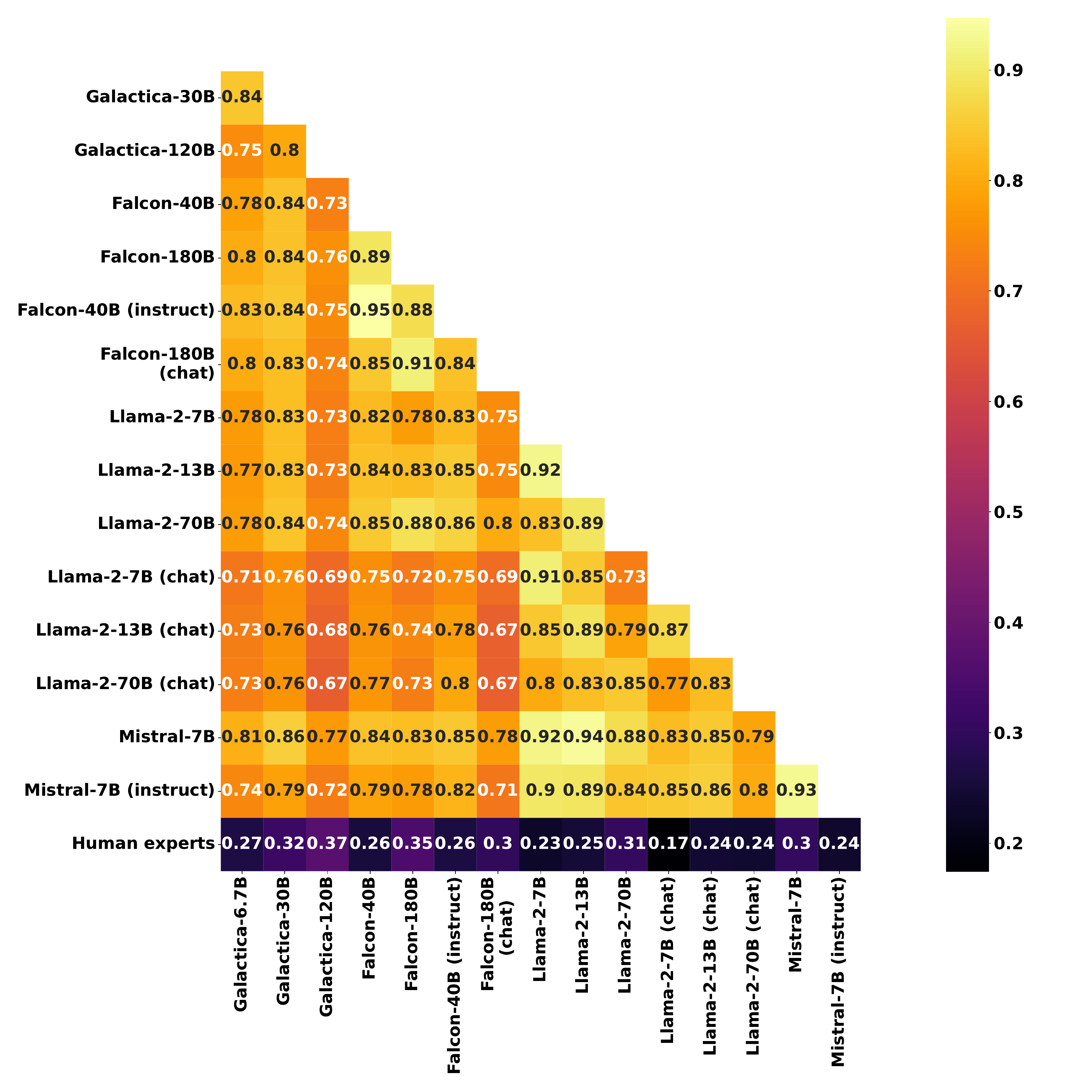}
    \caption{\textbf{Test difficulty correlation among LLMs and human experts (using GPT-4 created test cases).} Difference in perplexity between incorrect and correct abstracts is used to determine the relative difficulty of test cases. Spearman correlation is computed between ranked difficulties by LLMs and human experts. LLMs have an average Spearman correlation of $0.83$ ($\pm0.06$) whereas human experts have an average of $0.27$ ($\pm0.05$) with LLMs.}
    \label{fig:error_correlation_heatmap_machineCreated}
\end{figure}

\begin{figure}[H]
    \centering
\includegraphics[width=.8\textwidth]{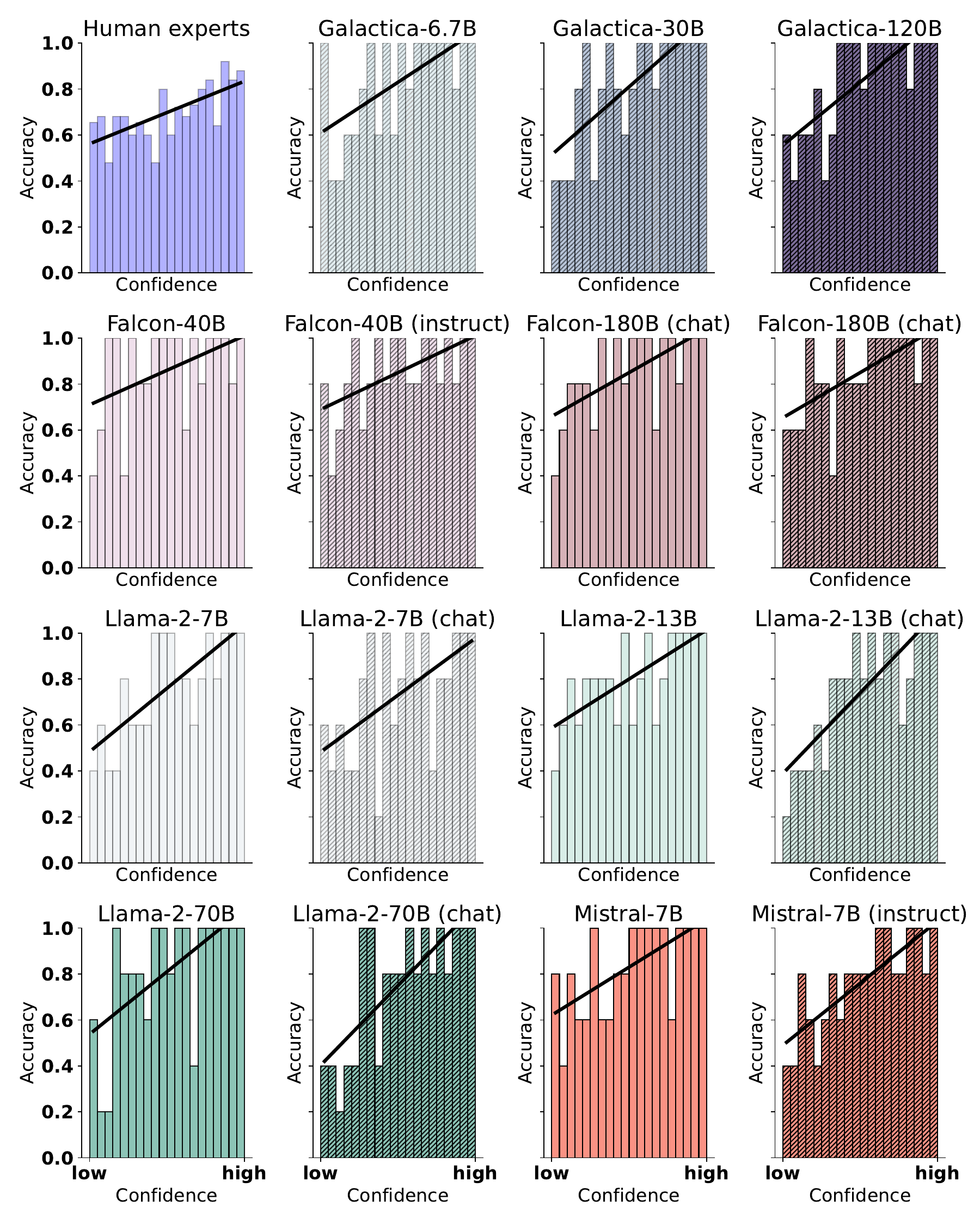}
    \caption{\textbf{Accuracy and confidence are calibrated for human experts and LLMs (using GPT-4 created test cases).} When human experts and LLMs are confident in their BrainBench judgments, they are more likely to be correct. Confidence ratings were sorted and placed in equally-sized bins with the mean accuracy for items in that bin plotted. The black regression line's positive slope for human experts and all LLMs indicates that confidence is well calibrated (i.e., higher confidence corresponds to higher accuracy).
    Calibration is beneficial for human-machine teams.}
    \label{fig:calibration_machineCreated}
\end{figure}

\begin{figure}[H]
    \centering
    \includegraphics[height=.7\textheight,width=.7\textwidth,keepaspectratio]{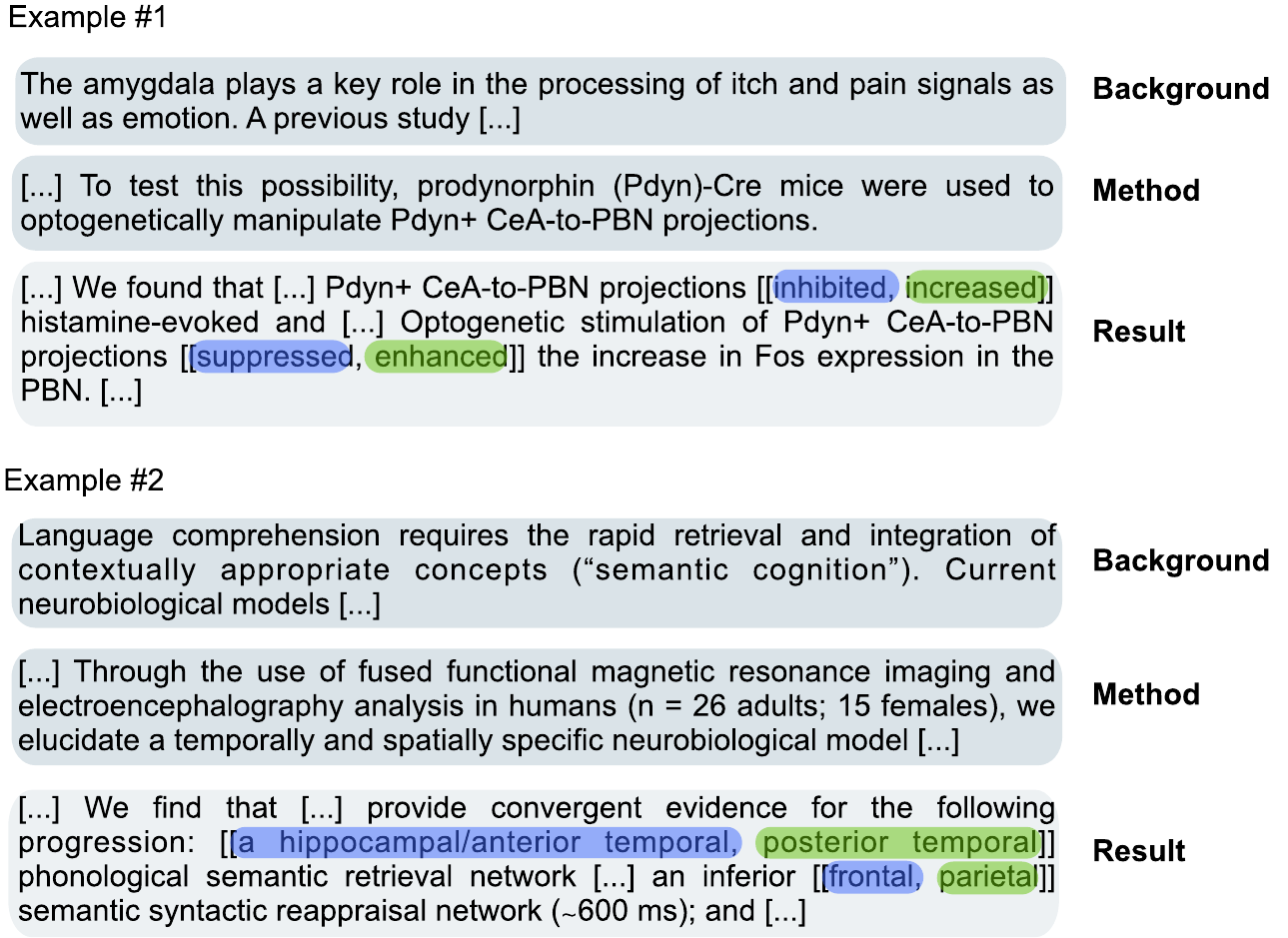}
    \caption{\textbf{BrainBench examples.} For a test case, the background and the method sections of the abstract are kept unchanged. Alternative findings are introduced by human experts or GPT-4. Participants choose from two options, the original and the altered, with the aim of selecting the actual (i.e., original) finding.}
    \label{fig:testcase_examples}
\end{figure}

\bibliography{references-brad,references-ken,references-ken-mend}
\bibliographystyle{Science}
\end{document}